\documentclass[12pt]{article}

\usepackage[dvips]{graphicx}
\usepackage{epsfig}
\usepackage{amsmath,amsfonts,amssymb,amsthm}
\usepackage{verbatim}
\usepackage{psfrag}
\usepackage{bm}
\usepackage{bbm}
\usepackage[square,comma,sort&compress,numbers]{natbib}
\usepackage{color}
\usepackage{slashed}

\usepackage{epsf,epsfig}
\usepackage{graphics}

\def\be{\begin{equation}}
\def\ee{\end{equation}}
\begin{document}
\thispagestyle{empty}

\begin{flushright}
{
\small
TTK-11-63\\
}
\end{flushright}

\vspace{0.4cm}
\begin{center}
\Large\bf\boldmath
Effective Theory of Resonant Leptogenesis in the
Closed-Time-Path Approach
\unboldmath
\end{center}

\renewcommand{\thefootnote}{\fnsymbol{footnote}}

\vspace{0.7cm}

\begin{center}
{Bj\"orn~Garbrecht and Matti Herranen\footnote[1]{Alexander-von-Humboldt
Fellow}}\\
\vskip0.3cm
{\it Institut f\"ur Theoretische Teilchenphysik und Kosmologie,\\ 
RWTH Aachen University, 52056 Aachen, Germany}\\
\end{center}
\vspace{.5cm}

\renewcommand{\thefootnote}{\arabic{footnote}}

\begin{abstract}
We describe mixing scalar particles and Majorana fermions using
Closed-Time-Path methods. From the Kadanoff-Baym equations,
we obtain the charge asymmetry, that is generated from
decays and inverse decays of the mixing particles. Within one
single formalism, we thereby treat Leptogenesis from oscillations and
recover as well the standard results for the asymmetry in
Resonant Leptogenesis, which apply when the oscillation
frequency is much larger than the decay rate. Analytic solutions
for two mixing neutral particles in a constant-temperature background
illustrate our results qualitatively. We also perform the modification
of the kinetic equations that is necessary in order to take account
of the expansion of the Universe and the washout of the asymmetry.
\end{abstract}

\newpage

\section{Introduction}

There are several different methods of calculating the lepton asymmetry,
that may be generated in the Early Universe. This variety of
approaches is a feature that Leptogenesis~\cite{Fukugita:1986hr} has in
common with many other topics in Particle Theory.

The approach that is
perhaps most commonly applied is to calculate the decay asymmetry
of individual right handed, singlet Majorana neutrinos $N$ from $S$-matrix
elements~\cite{Fukugita:1986hr,Luty:1992un,Covi:1996wh,Buchmuller:1997yu}.
These asymmetries, obtained by Quantum Field Theoretical methods,
are subsequently
substituted into classical Boltzmann equations, that describe the
macroscopic non-equilibrium
dynamics and the evolution of the asymmetry. By LSZ-reduction,
$S$-matrix elements can be obtained from time ordered $n$-point
Green functions.
We then enjoy the usual simplifications of time-ordered perturbation
theory, {\it i.e.} we can calculate the $n$-point functions using the standard diagrammatic
Feynman rules. The price to pay for this simplification
is cutting the direct link between the non-equilibrium
dynamics and the generation of the $CP$-asymmetry at the quantum level.
A particular problem turns out to be
due to the instability of the singlet neutrinos,
which causes that the $S$-matrix, that involves the $CP$-asymmetric decays and
inverse decays of these singlets, is not unitary, or rather, that it does not
correspond to an $S$-matrix in the proper sense.
As a consequence,
the unitary evolution of the non-equilibrium system must
be carefully
re-implemented, which is performed in practise
through the non-trivial method of
real intermediate state subtraction~\cite{Kolb:1979qa}. Without real
intermediate state subtraction, one would predict
from substituting the rates from the $S$-matrix 
elements into the network of Boltzmann equations the presence of a lepton
asymmetry even in thermal equilibrium, which
would be an unacceptable violation
of the $CPT$ theorem.

For Resonant
Leptogenesis~\cite{Flanz:1996fb,Covi:1996wh,Pilaftsis:1997dr,Pilaftsis:1997jf,Pilaftsis:2003gt,Pilaftsis:2005rv},
it is assumed that the mass difference
$\Delta M$ of two of the singlet neutrinos is small compared
to their average mass $\bar M$, or, more precisely, the energy
difference is small compared to the average energy.
Decays and inverse decays, that induce off-diagonal correlations of
the singlet neutrinos in their mass eigenbasis, then lead to a relatively large mixing,
which enhances $CP$-violation.
In the $S$-matrix approach, the mixing appears as a wave function
correction to the singlet neutrino propagator in the decay
diagrams~\cite{Covi:1996wh,IgnatievKuzminShaposhnikov,Botella:1990vf,Liu:1993tg}. The mixing that is calculated in this
way is time-independent, which excludes the possibility of 
incorporating the effect of oscillations.

Alternatively, one can avoid the subtleties concerning unitarity by
computing the real-time evolution of the quantum state from the outset.
This can either be achieved by using the canonical Hamiltonian approach or
functional, Closed-Time-Path (CTP)~\cite{Schwinger:1960qe,Keldysh:1964ud,Chou:1984es,Calzetta:1986cq} methods.
This may appear as a complicated task, but one should keep in
mind that the $CP$-asymmetry arises in the $S$-matrix language at the
one-loop level. In a real time approach, it turns out that
in a diagrammatic representation, we have to go to two-loop level.
However, the leading $CP$-violating effects can be extracted by imposing
some of the internal particles to be on-shell, which may be considered
as a generalisation of the Cutkosky rules~\cite{Cutkosky:1960sp}.
In the end, the integrals that are relevant for the leading order
prediction of the $CP$ asymmetry are identical to those encountered in the
$S$-matrix approach, but with the advantage that an overcounting of
particular processes in the Boltzmann equations,
that would violate unitarity, is avoided by construction.

The canonical
approach makes use of the property of the
Hamiltonian as the time evolution operator.
A blueprint of this canonical
approach, is the theory of neutrino
oscillations in
Refs.~\cite{Dolgov:1980cq,Barbieri:1990vx,Enqvist:1990ad,Enqvist:1991qj,Sigl:1992fn}. It has been
noticed that when supplemented
by $CP$-violation, the oscillations may lead to the creation of a lepton
asymmetry~\cite{Akhmedov:1998qx,Asaka:2005pn,Gagnon:2010kt}.
The possibility of describing mixing oscillations is one key advantage
of the canonical approach
when compared to using $S$-matrix elements. On the other
hand, without a set of diagrammatic rules, the perturbation expansion
is rather cumbersome. This is why the remaining important $CP$-violating
contribution, that is sometimes referred to as
direct $CP$-violation ({\it e.g.} in Ref.~\cite{Flanz:1996fb}), has not been calculated in
the Hamiltonian approach. In the $S$-matrix approach, this contribution
arises from the penguin-shaped vertex diagram.
Besides, the decays of the singlet neutrinos into Standard Model leptons
are described in practise by appealing again to the $S$-matrix formalism.
Therefore, one may state that within the Hamiltonian approach, no entirely
closed picture for the emergence of the lepton asymmetry has yet been developed.

In principle, it should be possible to derive a set of diagrammatic 
rules that simplifies the calculations within the Hamiltonian formalism.
However, such a set of rules
is readily available. It
arises in an elegant and rather intuitive manner from the functional
formulation of the
CTP formalism~\cite{Schwinger:1960qe,Keldysh:1964ud,Chou:1984es,Calzetta:1986cq}.
In the CTP approach, the Hamiltonian time evolution of the density matrix is
replaced by Schwinger-Dyson equations for the
two-point Green functions. These equations have four matrix components
from the two indices that denote the two branches of the CTP. The Kadanoff-Baym
equations are those of the Schwinger-Dyson equations, that encompass
the macroscopic evolution of the system.
Recently, this approach has proved successful in order to describe Leptogenesis
within a single theoretical framework~\cite{Buchmuller:2000nd,De Simone:2007rw,Garny:2009rv,Garny:2009qn,Anisimov:2010aq,Garny:2010nj,Beneke:2010wd,Beneke:2010dz,Garny:2010nz,Garbrecht:2010sz,Anisimov:2010dk}.
Both, $CP$-violation
from wave-function and vertex corrections are incorporated.
Unitarity issues are
resolved and an accurate account of all quantum-statistical effects on the
asymmetry is made. Moreover, the formulation in terms of Green functions
bears the potential of incorporating corrections from Thermal Field Theory
within this formalism. For the $CP$-conserving production rate of singlet 
neutrinos, several phenomenologically relevant thermal
corrections have recently been
calculated in Refs.~\cite{Asaka:2006rw,Kiessig:2010pr,Anisimov:2010gy,Salvio:2011sf,Kiessig:2011fw,Kiessig:2011ga,Laine:2011pq}.
When it comes to the rate of $CP$-violation, an aspect that is not dealt with
in Refs.~\cite{Buchmuller:2000nd,De Simone:2007rw,Garny:2009rv,Garny:2009qn,Anisimov:2010aq,Garny:2010nj,Beneke:2010wd,Beneke:2010dz,Garny:2010nz,Garbrecht:2010sz,Anisimov:2010dk} is the possibility of flavour oscillations between the
singlet neutrinos. The present work addresses this shortcoming.

While in Ref.~\cite{Beneke:2010wd}, the wave function correction is calculated using the
perturbative loop expansion, the present work captures the same effect by
solving the Kadanoff-Baym equations for the singlet neutrino propagator
directly. In the limit where the mass difference is much larger than
the decay rate $\Gamma_{\rm D}$ of the singlets, $\Delta M\gg\Gamma_{\rm D}$,
the standard perturbative result is reproduced. When
$\Delta M\sim\Gamma_{\rm D}$ or smaller, 
it becomes important that the
solution to the Kadanoff-Baym equations corresponds to a resummation of all wave-function insertions.
This is a generalisation of the
familiar procedure for time-independent situations, which can be performed
by the summation of a geometric series. The result obtained in the
present work from the
Kadanoff-Baym equations encompasses both, time-independent and oscillating 
contributions to the mixing of the singlet neutrinos.
We note that
for a model of the asymmetric decay of a scalar field, it was clarified
in Refs.~\cite{Liu:1993ds,Covi:1996fm} that
the Hamiltonian approach incorporates indeed the wave-function contribution
to the $CP$-asymmetry from decays in the vacuum
as it is usually calculated from $S$-matrix
elements.
The present work goes beyond this proof of
principle not only because it also treats the decays of Majorana fermions,
but also because it formulates the evolution of the asymmetry in an
interacting, finite-density system within the single framework of the
CTP formalism. Our results may therefore be employed for phenomenological studies
of Resonant Leptogenesis in the Early Universe.

In order to further describe the context of this present work, we note that
for computations of the asymmetry from $S$-matrix elements, it has
been realised that the resonant enhancement factor $1/\Delta M$ should
be regulated by the finite width of the singlet
neutrinos~\cite{Pilaftsis:1997dr,Pilaftsis:1997jf,Buchmuller:1997yu,Pilaftsis:2003gt,Pilaftsis:2005rv,Anisimov:2005hr}.
For solutions to the Kadanoff-Baym equations in Wigner space,\footnote{The Wigner transform is the Fourier transform of a
two-point function with respect to the relative coordinate,
while retaining the average coordinate, see
Appendix~\ref{appendix:CTP:Wigner}.}
it turns out that the
non-equilibrium distribution functions are not multiplied by a finite-width,
Breit-Wigner distribution, but rather by a zero width, Dirac
$\delta$-function. The observation of this somewhat curious property has lead to the 
speculation that these zero-width
solutions arise due to an incorrect treatment of pinch 
singularities that must be addressed by a non-perturbative
resummation~\cite{hep-ph/9802312,Greiner:1998ri}, that
is yet to be specified. This may be perceived as an obstacle for the formulation
of an effective theory of Resonant Leptogenesis within the CTP framework.
However, it was recently demonstrated analytically and numerically, that the vanishing width is a correct property of the out-of-equilibrium Wigner functions.
Within loop diagrams, the zero width is corrected for by a
resummation of derivative operators, which effectively shifts the argument of the
$\delta$-function from the real axis into the
imaginary direction of the complex plane~\cite{arXiv:1108.3688}.
In the Kadanoff-Baym equations, this effect may be of relevance when the tree-level scattering
processes are kinematically forbidden. As this is not the situation that we assume
for the present purposes, the corrections that are explained in Ref.~\cite{arXiv:1108.3688} are subdominant
in the final result for the asymmetry
and therefore are neglected here.
The results of Ref.~\cite{arXiv:1108.3688} imply however, that a straightforward
solution of the Kadanoff-Baym equations should indeed regulate the asymmetry
when $\Delta M\to 0$. Indeed, this is what we explicitly find in the present work.

After these comments, we present the outline of this paper:
In Section~\ref{section:scalar}, we use the
Schwinger-Dyson equations on the CTP, in particular the
Kadanoff-Baym equations, in order to calculate the asymmetry
that emerges from the decays of mixing neutral scalar particles.
The Kadanoff-Baym equations decompose into
constraint and kinetic equations. When $\Delta M\gg \Gamma_{\rm D}$,
one may use either of these equations in order to derive the asymmetry.
If this is not the case, one should use an expansion based on
$\Delta M\ll \bar M$. In this limit, the kinetic equations take
the form that is familiar from the time-evolution of a density matrix
in the Hamiltonian approach. In order to gain a qualitative 
understanding, we derive analytical solutions for the mixing
propagator in the case of two mixing flavours in a constant-temperature
background.

In Section~\ref{section:fermions}, we apply the same approximation
strategies
as for scalars to mixing Majorana neutrinos. As an
additional complication, we need to solve for the various spinor
components of the singlet neutrino propagator. However, it turns
out that the flavour dynamics shares the same essential features
with the scalar model. Even though our main interest is the
decay of Majorana fermions, a reader who is less interested
in these technical details can skip this Section and continue
with Section~\ref{section:illustrative}. There, we discuss
some features of the analytical solutions for the asymmetry in
a model in constant temperature background, for both, fermions
and scalars. In particular, we see that when
$\Delta M\gg \Gamma_{\rm D}$, the effect of oscillations of the singlet 
neutrinos averages out and the standard results for the decay
asymmetry from $S$-matrix elements apply. However, when
the condition
$\Delta M\gg \Gamma_{\rm D}$ does not apply, the oscillations
can have a significant effect on the asymmetry.

In Section~\ref{section:effective:theory}, we present the modifications
of the kinetic equations that arise in an expanding Universe. We 
therefore
eventually arrive at an effective theory for Resonant Leptogenesis,
that is formulated within the CTP framework.
In Section~\ref{section:conclusions}, there are a short summary of
this paper as well as concluding remarks.

\section{Scalar Model}
\label{section:scalar}

\subsection{CTP Approach to Generating a Charge Asymmetry from out-of-Equilibrium Decays}

The perhaps simplest model for generating a charge asymmetry from
out-of-equilibrium decays is given by~\cite{Covi:1996fm,Garny:2009rv,Garny:2009qn}
\begin{align}
{\cal L}=(\partial_\mu \varphi)(\partial^\mu \varphi^*)
-M^2_\varphi |\varphi|^2
+
\frac 12 (\partial_\mu \chi_i)(\partial^\mu \chi_i)
-\frac12 M^2_{\chi_{ij}} \chi_i\chi_j
-g_i\chi_i \varphi^2-g_i^*\chi_i {\varphi^*}^2
\,.
\end{align}
The field $\varphi$ is a complex scalar that can be associated
with a charge, while the $\chi_i$ are real singlet scalar fields, that
can create a charge
asymmetry within $\varphi$ when decaying out-of equilibrium.
We sum over repeated
indices $i,j$. We take here $i=1,2$, for simplicity.
The generalisation to more than two flavours of $\chi$ is
mostly straightforward, except for the
analytic formulae involving the diagonalisation of matrices that
we present in Section~\ref{section:constant:background}. We choose
a basis where the mass-matrix $M_\chi$ is diagonal.
The couplings $g_i$ are complex and of mass dimension {\it one}.

When replacing the $g_i$ by Yukawa couplings and
the complex scalar $\varphi$ by a Higgs and a lepton field --
which is straightforward to implement,
above model describes soft Leptogenesis from sneutrino
mixing~\cite{Grossman:2003jv,D'Ambrosio:2003wy,Fong:2011yx}.
This may be an interesting phenomenological application of
the methods developed in the present work.

In Wigner space, the Kadanoff-Baym equations for the Green functions
of the fields $\chi_i$ and $\varphi$ are
\begin{align}
\label{KB:allorders}
&\left[
k^2-\frac14 \partial^2_t +{\rm i} k^0\partial_t -M^2
\right]\Delta^{<,>}
-{\rm e}^{-{\rm i}\diamond}\{\Pi^H\}\{\Delta^{<,>}\}
-{\rm e}^{-{\rm i}\diamond}\{\Pi^{<,>}\}\{\Delta^{H}\}
\\\notag
&\hskip1cm =\frac12{\rm e}^{-{\rm i}\diamond}
\left(
\{\Pi^>\}\{\Delta^<\}
-\{\Pi^<\}\{\Delta^>\}
\right)
\end{align}
and the equations for the retarded and advanced propagators
\begin{align}
\label{retav:Wigner}
\left[
k^2+{\rm i}k^0\partial_t-\frac14 \partial_t^2 -M^2
\right]\Delta^{R,A}
-{\rm e}^{-{\rm i}\diamond}\{\Pi^H\}\{\Delta^{R,A}\}
\pm{\rm e}^{-{\rm i}\diamond}\{{\rm i}\Pi^{\cal A}\}
\{\Delta^{R,A}\}=1\,.
\end{align}
In these equations, $\Delta=\Delta(k,t)$ and $\Pi=\Pi(k,t)$,
and for the particular fields, we substitute
$\Delta\to\Delta_{\chi,\varphi}$, $\Pi\to\Pi_{\chi,\varphi}$
and $M\to M_{\chi,\varphi}$.
The derivation of these equations is discussed in detail in
Refs.~\cite{hep-ph/9802312,Prokopec:2003pj,Prokopec:2004ic,arXiv:1108.3688}.
As a quick reference, we list in Appendix~\ref{appendix:CTP:Wigner}
the definitions of the various Green functions and self energies
as well as of the $\diamond$ operator, that appears in Wigner
space. In order to keep the equations compact,
we often suppress the momentum and time arguments of
the Wigner functions in the following

Next, we keep only the zeroth order contributions in the operator $\diamond$,
which we justify in more detail below. By taking the
anti-hermitian and hermitian parts,
it is useful to split the Kadanoff-Baym equations into
the constraint,
\begin{align}
\label{constraint}
2\left[
k^2-\frac14 \partial^2_t
\right]\Delta^{<,>}
-\{M^2+\Pi^H,\Delta^{<,>}\}
-\{\Pi^{<,>},\Delta^{H}\}=
-\frac12[ {\rm i}\Pi^>, {\rm i}\Delta^<]
+\frac12[ {\rm i}\Pi^<, {\rm i}\Delta^>]
\,,
\end{align}
and the kinetic equations
\begin{align}
\label{kinetic}
2 {\rm i} k^0\partial_t
\Delta^{<,>}
-[M^2,\Delta^{<,>}]
=-\frac 12
\left(
\left\{{\rm i}\Pi^>,{\rm i}\Delta^<\right\}
-\left\{{\rm i}\Pi^<,{\rm i}\Delta^>\right\}
\right)
\,.
\end{align}
Of course, for the single-flavoured field $\varphi$, the
commutator involving $M^2_\varphi$ vanishes.

The self-energies for $\chi$ are given by
\begin{align}
{\rm i}\Pi^{ab}_{\chi_{ij}}(p)=
\int
\frac{d^4 k}{(2\pi)^4}
\left[
g_i g_j^*
{\rm i}\Delta_\varphi^{ab}(k){\rm i}\Delta_\varphi^{ab}(p-k)
+g_i^* g_j
{\rm i}\Delta_\varphi^{ba}(-k){\rm i}\Delta_\varphi^{ba}(k-p)
\right]
\end{align}
and for $\varphi$ by
\begin{align}
{\rm i}\Pi_\varphi^{ab}(p)
=\sum\limits_{ij}g_i g_j^*
\int\frac{d^4 k}{(2\pi)^4}
{\rm i}\Delta^{ab}_{\chi_{ij}}(k)
{\rm i}\Delta^{ab}_\varphi(p-k)\,.
\end{align}
We absorb $\Pi^H$ into a redefinition of $M$ and
can therefore effectively drop all terms involving it
henceforth.

Integration of the kinetic equations~(\ref{kinetic})
for $\varphi$ yields the evolution of the
charge asymmetry
\begin{align}
&\int\frac{dp^0}{2\pi}
{\rm i}p^0\partial_t {\rm i}\Delta_\varphi^{<,>}
=S_\varphi+W_\varphi
\\\notag
=&-\frac12
\int\frac{dp^0}{2\pi}
\left(
{\rm i}\Pi_\varphi^>(p){\rm i}\Delta^<_\varphi(p)
-{\rm i}\Pi_\varphi^<(p){\rm i}\Delta^>_\varphi(p)
\right)
\\\notag
=&-\frac12 g_i g_j^*
\int\frac{dp^0}{2\pi}\int\frac{d^4 k}{(2\pi)^4}
\left(
{\rm i}\Delta^>_{\chi_{ij}}(k) {\rm i}\Delta_\varphi^>(p-k)
{\rm i}\Delta^<_{\varphi}(p)
-
{\rm i}\Delta^<_{\chi_{ij}}(k) {\rm i}\Delta_\varphi^<(p-k)
{\rm i}\Delta^>_{\varphi}(p)
\right)\,.
\end{align}
In this form, the equation encompasses both, the washout $W_\varphi$ of
the charge of $\varphi$ as well as the source term for
the asymmetry $S_\varphi$.
When we assume that the charge density of $\varphi$ is small
compared to the number density of its quasi-particles, a
very useful simplification arises ({\it cf.} Ref~\cite{Beneke:2010wd}).
The washout term $W_\varphi$
can be approximated by including only the leading order effects, that
do not involve $CP$-violation. For the $CP$-violating
source $S_\varphi$, we may approximate the field $\varphi$ as
following an equilibrium
distribution. Decomposing
\begin{align}
{\rm i}\delta\Delta_{\chi_{ij}}
=
{\rm i}\Delta^{<,>}_{\chi_{ij}}
-{\rm i}\Delta^{<,>{\rm eq}}_{\chi_{ij}}
\end{align}
and using KMS relations~(\ref{KMS}), we then find for the
$CP$-violating source term
\begin{align}
\label{asymmetry:compact}
S_\varphi
=-g_i g_j^*
\int\frac{d^4q}{(2\pi)^4}
{\rm i}\delta\Delta_{\chi_{ij}}(q) \hat\Pi^{\cal A}_{\chi}(q)
\,.
\end{align}
where we define
\begin{align}
\hat\Pi^{\cal A}_\chi=\Pi^{\cal A}_{\chi_{ij}}/(g_i^* g_j+g_i g_j^*)\,.
\end{align}
The spectral self-energy  $\Pi^{\cal A}_\chi$ is a pivotal
ingredient to calculations of the charge asymmetry within the
CTP approach. In Appendix~\ref{appendix:selfergs}, the concrete
expression for its equilibrium form is presented. Note that
$\Pi^{\cal A}_{\chi_{ii}}(k)/k^0$ is the decay rate of $\chi_i$.

It remains to determine ${\rm i}\delta\Delta_{\chi_{ij}}(q)$,
the deviation of the field
$\chi$ from equilibrium. From the constraint equations,
we infer that the mass-diagonal components can be
written as
\begin{equation}
\label{ansatz:diag}
{\rm i}\delta\Delta_{\chi_{ii}}(p)
=2\pi \delta(p^2-M_{\chi_{ii}}^2)
\left(
\vartheta(p^0)\delta f_{\chi_{ii}}(\mathbf p)
+\vartheta(-p^0)\delta \bar f_{\chi_{ii}}(\mathbf p)
\right)
\,.
\end{equation}
The functions $\delta f_\chi$ and $\delta \bar f_\chi$ describe
the deviation of the quasi-particle distributions of the fields
$\chi$ from their equilibrium Bose-Einstein form. Just as the Wigner
functions, they are time-dependent, but we always suppress the time
argument.
For the diagonal components, the neutrality of the fields $\chi$,
Eq.~(\ref{neutrality:scalar}),
implies that
\begin{align}
\label{neutrality:scalar:f}
\delta f_{\chi_{ii}}(\mathbf p)=\delta \bar f_{\chi_{ii}}(\mathbf p).
\end{align}
Notice that the terms within the constraint equations that
involve $\Delta_\chi^H$ only contribute to the
solution for the equilibrium Green function, but not to
${\rm i}\delta\Delta_\chi$. For a single scalar flavour,
this matter is
explained in detail within Ref.~\cite{arXiv:1108.3688}.

It is useful to evaluate the kinetic and the constraint equations
in a distributional sense, by integrating over some
interval of $p^0$. For this purpose,
we extract the distribution functions from the propagator.
The neutrality condition~(\ref{neutrality:scalar:f}) must be
generalised for the off-diagonal components.
In order to keep the notation compact, we therefore define

\begin{align}
\label{def:fp}
\delta f_\chi(p)=\left\{
\begin{array}{l}
\delta f_\chi(\mathbf p)\;\;{\rm for}\;\;p^0>0\\
\delta \bar f_\chi(\mathbf p)\;\;{\rm for}\;\;p^0<0
\end{array}
\right.
\,,
\end{align}
{\it i.e.}  $\delta f_\chi(p)$
depends on $p^0$ only through ${\rm sign}(p^0)$.
We emphasise that this is just a convenient definition,
because the values of $\delta f_\chi(p)$ for far off-shell
momenta $p$ are irrelevant within our present approximations.
We then extract
these distribution functions through
\begin{align}
\label{eval:distributional}
\delta f_\chi(p^{\prime 0},\mathbf p)=
\int\limits_{{\cal I}_\pm}\frac{dp^0}{2\pi}
2p^0 {\rm sign}(p^0){\rm i}\delta\Delta_\chi(p)
\;\;\;\;
\forall p^{\prime 0}\;\Big|\;{\rm sign}(p^{\prime 0})=\pm
\,.
\end{align}
Here, ${\cal I}_\pm$ is a small interval around
the positive (negative) quasi-particle pole.
Note that the neutrality condition for scalar
fields~(\ref{neutrality:scalar}) and
hermiticity~(\ref{hermiticity:scalar})
imply that
\begin{align}
\label{neutrality:fchi}
\delta f_{\chi_{ij}}(p^0,\mathbf p)
=\delta f_{\chi_{ji}}(-p^0,\mathbf p)
=\delta f^*_{\chi_{ij}}(-p^0,\mathbf p)\,.
\end{align}

Integrating the kinetic
equations~(\ref{kinetic}) according to Eq.~(\ref{eval:distributional}),
we obtain
\begin{align}
\label{kinetic:flavour}
2{\rm i}k^0\partial_t\delta f_\chi(k)-[M^2_\chi,\delta f_\chi(k)]=
-{\rm i}\{\Pi^{\cal A}(k),\delta f_\chi(k)\}
\,.
\end{align}
These equations should be solved by setting $k^0$ to its value
at the quasi-particle pole. The solution is
$\delta f_\chi(\mathbf k)$ for $k^0>0$ and
$\delta \bar f_\chi(\mathbf k)$ for $k^0<0$. It
can then be completed according to Eq.~(\ref{def:fp}), keeping
in mind that $\delta f_\chi(k)$ is only physically
meaningful for momenta close to the quasi-particle poles.

The off-diagonal components of
${\rm i}\Delta_\chi$ can be determined based on two
separate approximations, depending on the parametric regime.
In order to define these regimes, we introduce
\begin{align}
\label{GammaD:scalar}
\Gamma_{\rm D}(p)\approx \underset{ij}{\max}
\left|\Pi^{\cal A}_{\chi_{ij}}(p)/p^0\right|
\end{align}
as a measure of the width of the singlet $\chi$ or, equivalently,
its decay rate. Besides, we define
\begin{subequations}
\label{DeltaMMBar:scalar}
\begin{align}
\Delta M&=|M_{\chi_{11}}-M_{\chi_{22}}|\,,
\\
\bar M&=\frac{M_{\chi_{11}}+M_{\chi_{22}}}{2}\,.
\end{align}
\end{subequations}
The two parametric regimes are given by
$\Delta M\gg\Gamma_{\rm D}$ and by $\Delta M\ll \bar M$.
When $\chi$ is weakly coupled, as it is a typical requirement
in scenarios for baryogenesis from out-of equilibrium
decays, we have $\Gamma_{\rm D}\ll \bar M$, such that there
is an overlap of both regimes when
$\Gamma_{\rm D}\ll\Delta M\ll \bar M$. In that situation,
both approximations should lead to the same predictions
for the asymmetry.

The particular approximation strategies are now as
follows:
\begin{itemize}
\item
When $\Delta M\gg\Gamma_{\rm D}$, we can
calculate the off-diagonal components of
${\rm i}\delta\Delta_\chi$ as higher-order corrections
to the diagonal ones. For simplicity, we
assume that only $\chi_i$ deviates from equilibrium.
The general case follows from the superposition of
the procedure that we describe here.
In this approximation, the off-diagonal components share
the diagonal mass-shells, {\it i.e.} the poles
of these quasi-particles are given by
\begin{align}
\label{ansatz:offdiag}
{\rm i}\delta\Delta_{{\chi}_{ij}}(p)
=2\pi\delta(p^2-M_{\chi_{ii}}^2)\delta f_{\chi_{ij}}(p)\,,\qquad
{\rm i}\delta\Delta_{{\chi}_{ji}}(p)
=2\pi\delta(p^2-M_{\chi_{ii}}^2)\delta f_{\chi_{ji}}(p)
\\\notag
\qquad\textnormal{for}\;\;\Gamma_{\rm D}\ll \Delta M
\,.
\end{align}
As we show in Section~\ref{section:scalar:nondeg},
using either the kinetic or the constraint equations,
it indeed follows that
the off-diagonal components of the propagator are suppressed
by factors of order $\Gamma_{\rm D}/\Delta M$ compared to
the diagonal ones. These off-diagonal components then enter
the source term $S_\varphi$ as given in Eq.~(\ref{asymmetry:compact}),
while higher order corrections to this are again suppressed by
factors $\sim\Gamma_{\rm D}/\Delta M$.
In principle, the same approximation strategy is employed
in Ref.~\cite{Garny:2009qn}.
\item
When  $\Delta M\ll\bar M$, we have no procedure yet
in order to determine the precise location of the quasi-particle poles,
but know that these are located at $p^2-\bar M^2$, up to a
relative error of order $\Delta M/\bar M$.
We can therefore approximate the off-diagonal
out-of-equilibrium Wightman function as
\begin{align}
\label{OS:scalar:resonant}
{\rm i}\delta\Delta_{\chi_{ij}}(p)
=2\pi\delta(p^2-\bar M^2)
\delta f_{\chi_{ij}}(p)
\qquad\textnormal{for}\;\;\Delta M\ll \bar M
\,.
\end{align}
When using this in order to solve
the kinetic equation~(\ref{kinetic:flavour}) as described
above, we expect a
relative error of order $\Delta M/\bar M$.
Note that this approximation strategy is similar to
the one that is employed in Refs.~\cite{Cirigliano:2009yt,Cirigliano:2011di}.
\end{itemize}
In both cases, the result for ${\rm i}\Delta_{\chi_{ij}}$ can then
be substituted into the source term~(\ref{asymmetry:compact}).

It is now the point to come back to the truncation of higher
orders of the
$\diamond$ operator, a procedure known as the gradient
expansion.
It is
justified, provided we may neglect contributions to the finite width
of the propagators,
which are of relevance when the reaction $\chi\leftrightarrow 2\varphi$
is kinematically forbidden~\cite{arXiv:1108.3688}. Furthermore,
contributions to the gradient expansion from fast flavour oscillations,
which are systematically treated in Refs.~\cite{HKR1,HKR2,HKR3,Glasgow,Thesis_Matti,HKR4,FHKR1,HKR5},
must not
give sizable corrections to the phenomenological results. In the
resonant regime, such corrections are suppressed by the small
oscillation frequency and therefore
effectively by factors of order $\Delta M/\bar M$,
while in the non-resonant regime, the suppression results from the
averaging of the oscillatory contribution to the source for the 
asymmetry, as we discuss below.
Besides these considerations, the higher order
terms in the gradient expansion yield contributions of order
$H/T\sim T/m_{\rm Pl}$, where $H$ is the Hubble expansion rate,
$m_{\rm Pl}$ the Planck mass and $T$ the temperature
of the Universe,
which sets the typical energy of a quasi-particle. For temperatures
sufficiently below the Planck scale, corrections of this type
should therefore be negligible.

\subsection{The Regime $\Delta M\gg \Gamma_{\rm D}$}
\label{section:scalar:nondeg}

In this regime, when using the ansatz~(\ref{ansatz:offdiag}),
we find the following approximate solution
to the kinetic equations~(\ref{kinetic:flavour}):
\begin{subequations}
\label{deltaf:nonden}
\begin{align}
\delta f_{\chi_{ii}}(k)&=\delta f^0_{\chi_{ii}}(k){\rm e}^{-\frac{\Pi_{\chi{ii}}^{\cal A}(k)}{k^0} t}\,,
\\
\label{kinetic:hierarchical:offdiag}
\delta f_{\chi_{ij}}(k)&={\rm i}\frac{\Pi^{\cal A}_{\chi_{ij}}(k)}{M_{\chi_{ii}}^2-M_{\chi_{jj}}^2}
\delta f_{\chi_{ii}}(k)=-\delta f_{\chi_{ji}}(k)\,.
\end{align}
\end{subequations}
We assume here that only $\chi_i$ is excited by deviating
from its equilibrium distribution, {\it i.e.}
$\delta f^0_{\chi_{ii}}=\delta f_{\chi_{ii}}\big|_{t=0}$,
$\delta f_{\chi_{jj}}\big|_{t=0}=0$ and
we
set $k^0=\pm\sqrt{\mathbf k^2+M^2_{\chi_{ii}}}$.
The case with an excited $\chi_j$ can be written down in analogous manner, taking 
account of the different underlying mass shell. As anticipated,
the off-diagonal components are suppressed by a factor of order
$\Gamma_{\rm D}/\Delta M$.

When substituting $\delta f_{\chi_{ij}}$ into the Wightman function
according to Eq.~(\ref{ansatz:offdiag}) and eventually
into the $CP$-violating source term~(\ref{asymmetry:compact}) for $\varphi$,
we recover the known answer for $\Delta M\gg \Gamma_{\rm D}$
and in finite density background:
\begin{align}
\label{source:phi:hierarch}
S_\varphi&=
-{\rm i}
\int\frac{d^4 q}{(2\pi)^4} 2\pi\delta(q^2-M_{\chi_{ii}}^2)
(g_i g_j^*-g_i^* g_j)\frac{g_i g_j^*+g_i^* g_j}{M_{\chi_{ii}}^2-M_{\chi_{jj}}^2}
(\hat\Pi_\chi^{\cal A}(q))^2
\delta f_{\chi_{ii}}(q)
\\\notag
&=
-2{\rm i}
(g_i^2{g_j^*}^2-{g_i^*}^2g_j^2)
\int\frac{d^3 q}{(2\pi)^3 2 \sqrt{\mathbf q^2+M_{\chi_{ii}}^2}}
\frac{(\hat\Pi^{\cal A}_{\chi}(q))^2}{M_{\chi_{ii}}^2-M_{\chi_{jj}}^2}
\delta f_{\chi_{ii}}(\mathbf q)\,,
\end{align}
{\it cf.} Ref.~\cite{Garny:2009qn}.

Taking account of the time derivative~(\ref{kinetic:flavour}) and of the
anticommutator terms that involve the off-diagonal components
of $f_\chi$, we obtain the leading finite-width corrections that
appear within the denominator:
\begin{align}
\delta f_{\chi_{ij}}(k)&={\rm i}
\frac{
\Pi^{\cal A}_{\chi_{ij}}(k)
}
{
M_{\chi_{ii}}^2-M_{\chi_{jj}}^2
-{\rm i}\Pi^{\cal A}_{\chi_{ii}}+{\rm i}\Pi^{\cal A}_{\chi_{jj}}
}
\delta f_{\chi_{ii}}(k)=-\delta f_{\chi_{ji}}(k)\,.
\end{align}
Notice however, that this result still relies on the assumption
$\Delta M\gg\Gamma_{\rm D}$.

The same result can be obtained when using
the constraint equations~(\ref{constraint}).
For given ${\rm i}\delta \Delta_{ii}$ (and assuming that
${\rm i}\delta \Delta_{ij}=0$ for $i\not=j$ at zeroth
order in $\Gamma_{\rm D}/\Delta M$), the off-diagonal
components follow from
\begin{align}
\left[2k^2-(M^2_{\chi_{ii}}+M^2_{\chi_{jj}})\right]{\rm i}\delta \Delta_{ij}
={\rm i}\Pi^{\cal A}_{\chi_{ij}} {\rm i}\delta\Delta_{\chi_{ii}}
+\left({\rm i}\Pi^{\cal A}_{\chi_{ii}}-{\rm i}\Pi^{\cal A}_{\chi_{jj}}\right)
{\rm i}\delta \Delta_{ij}
\,.
\end{align}
The crucial term on the right hand side arises from the
commutator of the spectral self-energy and the distribution function
and is neglected in earlier literature.
Notice that the induced off-diagonal correlation is not oscillating.
This is in contrast to earlier approximations, where the
right hand side of the constraint equations is neglected~\cite{HKR1,HKR2,HKR3,Glasgow,Thesis_Matti,HKR4,FHKR1,HKR5,Konstandin:2005cd}.
In conjunction with Eqs.~(\ref{ansatz:diag})
and~(\ref{ansatz:offdiag}), in particular
with the on-shell condition
$k^2=M^2_{\chi_{ii}}$, we find consistency with the
solution~(\ref{kinetic:hierarchical:offdiag}).

\subsection{The Regime $\Delta M \ll\bar M$ and Damped $2\times 2$ Flavour Oscillations in a Constant Background}
\label{section:constant:background}

In this regime, the calculation should account for the potential
impact of flavour oscillations. Scalar and fermionic fields
share the same basic aspects of flavour dynamics.
For the purpose of generalisation, we therefore
recast Eq.~(\ref{kinetic:flavour})
as
\begin{align}
\label{flavour:dynamic:general}
\partial_t\delta f
+\frac{\rm i}2[\Omega,\delta f]=
-\frac12\{\Gamma,\delta f\}\,,
\end{align}
where
\begin{align}
\label{OmegaGamma:Scalar}
\Omega=M_\chi^2/k^0\,,\qquad \Gamma=\Pi_\chi^{\cal A}/k^0
\end{align}
and $\delta f=\delta f_\chi$. Of course, we can identify
this with the evolution equation for a density matrix in the
Hamiltonian formalism~\cite{Dolgov:1980cq,Barbieri:1990vx,Sigl:1992fn,Akhmedov:1998qx,Asaka:2005pn,Gagnon:2010kt}.

While this is not the case in the Early Universe, it is nonetheless
instructive to consider the situation when $\Omega$ and $\Gamma$
are time-independent.
The solution to equation~(\ref{flavour:dynamic:general}) is
then given by
\begin{align}
\label{sol:const:bg:formal}
\delta f(t)={\rm e}^{-\frac{\rm i}2\Omega t-\frac12\Gamma t}
\delta f(t=0)
{\rm e}^{\frac{\rm i}2\Omega t-\frac12 \Gamma t}\,.
\end{align}
Since $M_\chi$ is diagonal, we have chosen to work in
a flavour basis where $\Omega$ is diagonal as well. In this basis,
$\Gamma$ is non-diagonal in general, which is crucial in order to obtain $CP$-violating effects.
Notice that $\Omega$ and $\Gamma$ are both hermitian.
Let the matrix
\begin{align}
\Xi=\Omega-{\rm i}\Gamma
\end{align}
be diagonalised by the transformation
\begin{align}
\Xi_{\rm D}=U\Xi U^{-1}\,.
\end{align}
We choose the parametrisation
(see {\it e.g.} Ref.~\cite{physics/0607103})
\begin{align}
U=
\left(
\begin{array}{cc}
c & t_1 c\\
-t_2 c & c
\end{array}
\right)\,,
\end{align}
where
\begin{align}
\label{diagonalisation:parameters}
\begin{array}{rclrcl}
\Delta&=&\frac12(\Xi_{11}-\Xi_{22})\,,\qquad&
D&=&{\rm sign}({\rm Re}[\Delta])\sqrt{\Delta^2+\Xi_{12}\Xi_{21}}\,,
\\
t_1&=&\frac{\Xi_{12}}{\Delta+D}\,,&
t_2&=&\frac{\Xi_{21}}{\Delta+D}\,,
\\
c&=&\frac{1}{\sqrt{1+t_1t_2}}\,,&
\delta&=&\frac{\Xi_{12}\Xi_{21}}{\Delta+D}
\,,\\
\Xi_{{\rm D}11}&=&\Xi_{11}+\delta\,,&
\Xi_{{\rm D}22}&=&\Xi_{22}-\delta\,.
\end{array}
\end{align}
The matrix
\begin{align}
\Xi^{\rm c}=\Omega+{\rm i}\Gamma
\end{align}
is diagonalised by
\begin{align}
\Xi^{\rm c}_{\rm D}=V\Xi^{\rm c} V^{-1}\,,
\end{align}
where $V$ can be constructed in the same way as $U$. Notice that
$\Xi^{\rm c}_{\rm D}=\Xi_{\rm D}^*$, $U^\dagger=V^{-1}$ and $\det U=1$.

We can thus express
\begin{align}
\label{toy:oscillations}
\delta f(t)=U^{-1}{\rm e}^{-{\rm i}\Xi_{\rm D}t} U
\delta f(0)
V^{-1}{\rm e}^{{\rm i}\Xi^{\rm c}_{\rm D}t}V
=U^{-1}{\rm e}^{-{\rm i}\Xi_{\rm D}t} U
\delta f(0)
U^\dagger{\rm e}^{{\rm i}\Xi^{\rm c}_{\rm D}t}{U^\dagger}^{-1}\,.
\end{align}
Note that $\delta f(t)$ is hermitian by construction, provided
$\delta f(0)$ is.
From this formula, we can infer the distribution functions
of $\chi$ and their correlations
as a function of time in a
straightforward way.

For this purpose, appropriate boundary conditions for
$\delta f$ need to be specified. In order to choose these,
it is useful to first consider the equilibrium propagators.
We express these various
propagators in terms of their adjugate matrices
\begin{subequations}
\label{scalar:finite-width}
\begin{align}
\Delta_\chi^R&=
\frac1{{k^0}^2-\omega^2+{\rm i}\Pi_\chi^{\cal A}}
=\frac1{\cal D}{\rm adj}({k^0}^2-\omega^2+{\rm i}\Pi_\chi^{\cal A})\,,
\\
\Delta_\chi^A&=
\frac1{{k^0}^2-\omega^2-{\rm i}\Pi_\chi^{\cal A}}
=\frac1{{\cal D}^*}{\rm adj}({k^0}^2-\omega^2-{\rm i}\Pi_\chi^{\cal A})\,,
\\
\Delta_\chi^{\cal A}&=
\frac{\rm i}{2}(\Delta_\chi^R-\Delta_\chi^A)
=\frac{\rm i}{2|{\cal D}|^2}
\left[
{\rm adj}({k^0}^2-\omega^2)
({\cal D}-{\cal D}^*)
+{\rm i}\,{\rm adj}(\Pi_\chi^{\cal A})
({\cal D}+{\cal D}^*)
\right]
\,,
\\
\Delta_\chi^{H}&=
\frac{1}{2}(\Delta_\chi^R+\Delta_\chi^A)
=\frac{\rm i}{2|{\cal D}|^2}
\left[
{\rm adj}({k^0}^2-\omega^2)
({\cal D}+{\cal D}^*)
+{\rm i}\,{\rm adj}(\Pi_\chi^{\cal A})
({\cal D}-{\cal D}^*)
\right]
\,,
\end{align}
\end{subequations}
where
\begin{align}
{\cal D}=\det\left[{k^0}^2-\omega^2+{\rm i}\Pi_\chi^{\cal A}\right]\,,
\end{align}
and $\omega^2=\mathbf k^2+M^2_\chi$.
These solutions can be obtained from the equations for the retarded and advanced
propagators~(\ref{retav:Wigner}). Besides,
the equilibrium Wigner functions
${\rm i}\Delta_\chi^{<{\rm eq}}(k)=2\Delta_\chi^{\cal A}(k)/(\exp(\beta k^0)-1)$
and ${\rm i}\Delta_\chi^{>{\rm eq}}(k)=2\Delta_\chi^{\cal A}(k)[1+1/(\exp(\beta k^0)-1)]$
can be consistently obtained as
solutions to the
constraint equations~(\ref{constraint}) as well,
{\it cf.} Ref.~\cite{arXiv:1108.3688} for a detailed discussion. 

As we assume a mass-diagonal basis, where $\omega^2$ is diagonal,
it becomes clear that in the hierarchical limit, this is indeed
the appropriate basis for defining the density of quasi-particles.
In particular, notice that for
$\Delta M_{\chi_{ij}}\gg\Gamma$, it follows
\begin{align}
{\rm i}\Delta^{\cal A}_{\chi_{ii}}\approx
\frac{\Pi^{\cal A}_{\chi_{ii}}}{({k^0}^2-\omega_{ii}^2)^2+{\Pi_{\chi_{ii}}^{\cal A}}^2}
\,,
\end{align}
while
\begin{align}
{\rm i}\Delta^{\cal A}_{\chi_{ij}}\approx
\Pi^{\cal A}_{\chi_{ij}}\left(\frac{1}{({k^0}^2-\omega_{ii}^2)^2+{\Pi_{\chi_{ii}}^{\cal A}}^2}
\frac{{k^0}^2-\omega_{ii}^2}{{k^0}^2-\omega_{jj}^2}
+
\frac{1}{({k^0}^2-\omega_{jj}^2)^2+{\Pi_{\chi_{jj}}^{\cal A}}^2}
\frac{{k^0}^2-\omega_{jj}^2}{{k^0}^2-\omega_{ii}^2}
\right)
\,,
\end{align}
where $\omega_{ii}^2=\mathbf k^2+M_{\chi_{ii}}^2$.
Hence, the diagonal contributions are of the Breit-Wigner type,
as required by the spectral sum rule
\begin{align}
\int\frac{d k^0}{2\pi}k^0\Delta^{\cal A}_{ij}
=\delta_{ij}\,,
\end{align}
while the off-diagonal entries are representations of the
principal value that cannot be associated with particle number.

Therefore, while it is tempting to choose for
Eq.~(\ref{toy:oscillations}) initial conditions where
$ U\delta f(0)V^{-1}$ is a diagonal matrix,
because then $\delta f(t)$ has no oscillatory contributions,
we should instead choose $\delta f(0)$ diagonal in order
to describe a state that initially
exhibits no mixing of the mass eigenstates.
For illustrative purposes, we therefore
consider a situation where $\chi_1$ is out-of-equilibrium, while $\chi_2$ is in equilibrium,
{\it i.e.}
\begin{align}
\delta f(0)=
\left(
\begin{array}{cc}
\delta f_0 & 0\\
0 & 0
\end{array}
\right)\,.
\end{align}
Eq.~(\ref{toy:oscillations}) then yields
\begin{subequations}
\label{densities:analytical}
\begin{align}
\delta f_{11}(t)&=
\delta f_0\frac{
\left(
(\Delta^*+D^*)^2{\rm e}^{\frac{\rm i}2 \Xi^{\rm c}_{{\rm}D11} t}
-|\Gamma_{12}|^2{\rm e}^{\frac{\rm i}2 \Xi^{\rm c}_{{\rm}D22} t}
\right)
\left(
(\Delta+D)^2{\rm e}^{-\frac{\rm i}2 \Xi_{{\rm}D11} t}
-|\Gamma_{12}|^2{\rm e}^{-\frac{\rm i}2 \Xi_{{\rm}D22} t}
\right)
}
{\left|(\Delta+D)^2-|\Gamma_{12}|^2\right|^2}
\,,
\\
\label{offdiag:analytical}
\delta f_{12}(t)
&=\delta f_0\frac{
{\rm i}\Gamma_{12}(\Delta^*+D^*)
\left(
{\rm e}^{\frac{\rm i}2 \Xi^{\rm c}_{{\rm}D11} t}
-{\rm e}^{\frac{\rm i}2 \Xi^{\rm c}_{{\rm}D22} t}
\right)
\left(
(\Delta+D)^2{\rm e}^{-\frac{\rm i}2 \Xi_{{\rm}D11} t}
-|\Gamma_{12}|^2{\rm e}^{-\frac{\rm i}2 \Xi_{{\rm}D22} t}
\right)
}
{\left|(\Delta+D)^2-|\Gamma_{12}|^2\right|^2}
\,,
\\
\delta f_{21}(t)&=\delta f_{12}^*(t)
\,,
\\
\delta f_{22}(t)
&=\delta f_0\frac{
|\Gamma_{12}|^2|\Delta+D|^2
\left(
{\rm e}^{\frac{\rm i}2 \Xi^{\rm c}_{{\rm}D11} t}
-{\rm e}^{\frac{\rm i}2 \Xi^{\rm c}_{{\rm}D22} t}
\right)
\left(
{\rm e}^{-\frac{\rm i}2 \Xi_{{\rm}D11} t}
-{\rm e}^{-\frac{\rm i}2 \Xi_{{\rm}D22} t}
\right)
}
{\left|(\Delta+D)^2-|\Gamma_{12}|^2\right|^2}
\,.
\end{align}
\end{subequations}
In the $CP$-violating source term for the scalar model~(\ref{asymmetry:compact}),
this enters when using~Eq.~(\ref{OS:scalar:resonant}).

It is interesting to study the behaviour of the
solutions~(\ref{sol:const:bg:formal}) and~(\ref{densities:analytical})
in the degenerate limit, where $M_{\chi_{11}}=M_{\chi_{22}}$.
Because $\Pi^{\cal A}_{\chi}$ is real symmetric, $\Gamma_{12}$
is real. We can then see from Eq.~(\ref{asymmetry:compact}), that
the $CP$-violating source $S_\varphi$ is proportional to
the imaginary part of $\delta f_{12}$. In
Eq.~(\ref{sol:const:bg:formal}), $\Omega$ commutes with all terms
for $\Delta M=0$, since it is proportional to the
identity matrix. Therefore, no contribution to $S_\varphi$ arises
in that situation. From Eq.~(\ref{offdiag:analytical}), we arrive at
the same conclusion, because both $\Delta$ and $D$ are purely imaginary
for $\Delta M=0$
and ${\rm Re}[\Xi_{{\rm D}11}]={\rm Re}[\Xi_{{\rm D}22}]$.
Therefore, $\delta f_{12}(t)$ turns out to be real, which implies
that $S_\varphi=0$.

We should also comment on the case where $(\Delta+D)^2-|\Gamma_{12}|^2=0$.
While from  Eq.~(\ref{sol:const:bg:formal}),
it is immediately clear that the result for $\delta f(t)$ must be
non-singular also in this limit, we need to explicitly convince ourselves
of the cancellation of singularities in Eqs.~(\ref{densities:analytical}).
From Eqs.~(\ref{diagonalisation:parameters}), we find that
$\Delta^2=\Xi_{12}\Xi_{21}=|\Gamma_{12}|^2$, $D=0$
and $\Xi_{{\rm D}11}=\Xi_{{\rm D}22}=\Xi_{11}-\sqrt{-\Xi_{12}\Xi_{21}}$
in that situation. Substitution into Eqs.~(\ref{densities:analytical})
then explicitly reveals the cancellation. In conclusion, we
find $S_\varphi=0$ for $\Delta M=0$, even when the fields $\chi_{1,2}$
are distinguishable through their interactions $g_{1,2}$ with
the field $\varphi$. Of course, one can alternatively arrive at
the same conclusion through a field redefinition through
a unitary transformation of $\chi$, that may remove the phases
from the couplings $g$ but leaves $M_\chi$ invariant for
$\Delta M=0$.

Finally,
for $\Delta M\gg \Gamma_{\rm D}$,
we can cross check the solutions~(\ref{offdiag:analytical}) by
neglecting the oscillatory contributions based on the assumption that these average out in the
final asymmetry, as we discuss in Section~\ref{section:illustrative}.
In that case, we find
\begin{align}
\delta f_{12}=\frac{{\rm i}\Pi^{\cal A}_{12}}{M_{\chi_{11}}^2-M_{\chi_{22}}^2}
\delta f_0{\rm e}^{-\frac{\Pi^{\cal A}_{11}}{k^0} t}\,,
\end{align}
which is in agreement with the result~(\ref{deltaf:nonden}), where the approximation of
non-degeneracy has been applied at an earlier stage of the calculation.

\section{Fermions}
\label{section:fermions}

\subsection{CTP Approach to Leptogenesis}
\label{section:CTPLepto}

For our treatment of Leptogenesis, we follow Ref.~\cite{Beneke:2010wd}, where
the CTP approach in the non-resonant regime is formulated.
A lot of that material applies to the resonant regime
as well, which
is the main objective of the present work.
Besides, we employ the results of Ref.~\cite{Beneke:2010wd} as a benchmark
in order
to test the results that are derived here.
Because they are applicable
arbitrarily close to the limit of mass-degenerate singlet
neutrinos, the present results are a generalisation of those
presented in Ref.~\cite{Beneke:2010wd}, and we show that
both agree
sufficiently far away from the mass degeneracy. Notice that
we do not reiterate the calculation of the asymmetry from the
vertex (penguin) correction in the present paper.

The Lagrangian that gives rise to the standard scenario of Leptogenesis
in the unflavoured regime is
\begin{align}
\label{Lagrangian}
{\cal L}=\frac{1}{2}\bar\psi_{Ni}({\rm i} \partial\!\!\!/-M_{ij}) \psi_{Nj}
+\bar\psi_\ell{\rm i}\partial\!\!\!/\psi_\ell
+(\partial^\mu\phi^\dagger)(\partial_\mu \phi)
-Y_i^*\bar\psi_\ell \phi^\dagger P_{\rm R}\psi_{Ni}
-Y_i\bar\psi_{Ni}P_{\rm L}\phi\psi_\ell\,.
\end{align}
Here, $N_i$ are the singlet right-handed Majorana neutrinos, that
correspond to the fields $\chi_i$ in the scalar model. Their masses
are given by the mass matrix $M$, for which we choose a diagonal basis,
unless explicitly stated otherwise.
Again, we take $i=1,2$, for simplicity, while a generalisation
to more flavours of $N$ is straightforward. The Standard Model
Higgs doublet field is $\phi$, and $\ell$ is a linear combination of
the three left-handed lepton doublets of the
Standard Model. $\psi_\ell$ and $\psi_N$ are the spinors
associated with $\ell$ and $N$.
Note that the diagonal form of the Yukawa couplings
$Y$, that is assumed in the Lagrangian~(\ref{Lagrangian}), can be
achieved by a unitary transformation of the left-handed
lepton flavours.
The ${\rm SU}(2)_{\rm L}$ indices are contracted in a gauge-invariant way, 
{\it i.e.} $\phi\psi_\ell=-\phi_A\epsilon_{AB}\psi_{\ell B}$
and $\bar\psi_{\ell}\phi^\dagger=(\bar\psi_{\ell})_A \epsilon_{AB}
(\phi^\dagger)_B$, where $\epsilon_{AB}$ is the antisymmetric $2\times 2$ 
tensor with $\epsilon_{12}=1$. The four-component spinor $\psi_{Ni}$
observes the Majorana constraint~(\ref{Majorana:condition}).
The chiral projectors are
$P_{{\rm L},{\rm R}}=(1\mp\gamma_5)/2$.

In the Lagrangian of the form~(\ref{Lagrangian}), $M$ may be a general
complex symmetric matrix. Through field redefinitions of $N$, it is possible to
bring $M$ to a real diagonal form, along with an according reparametrisation
of $Y_i$. Yet, once a basis is chosen where $M$ is real and
diagonal in the vacuum, thermal corrections may lead to a real symmetric
effective mass that adds to $M$. The off-diagonal components can then be
be removed through an addtional orthogonal transformation, but in general not
in a temperature-independent way. Yet, choosing a basis where $M$ is real and symmetric,
has the advantage that the $CP$-violating phases are
isolated within $Y_i$. Moreover, the separation of the Kadanoff-Baym equations into kinetic
and constraint equations by taking the hermitian and anti-hermitian parts
is facilitated when $M$ is purely hermitian. For these reasons, we continue
the discussion under the assumption that $M$ is real and symmetric.

The propagators for the left-handed lepton and the Higgs
field are listed in Appendix~\ref{appendix:propagatorshiggslepton}.
At zeroth order in perturbation theory,
the diagonal components of the singlet neutrino
propagators are parametrised in terms of the distribution functions
$f_N$ as
\begin{subequations}
\label{prop:N:expl}
\begin{align}
{\rm i}S_{N{ii}}^{<}(p)
&=-2\pi\delta(p^2-M_{ii}^2)(p\!\!\!/+M_{ii})\left[
\vartheta(p_0)f_{N{ii}}(\mathbf{p})
-\vartheta(-p_0)(1-f_{N{ii}}(-\mathbf{p}))
\right]\,,\\
{\rm i}S_{N{ii}}^{>}(p)
&=-2\pi\delta(p^2-M_{ii}^2)(p\!\!\!/+M_{ii})\left[
-\vartheta(p_0)(1-f_{N{ii}}(\mathbf{p}))
+\vartheta(-p_0)f_{N{ii}}(-\mathbf{p})
\right]\,,\\
{\rm i}S_{N{ii}}^{T}(p)
&=
\frac{{\rm i}(p\!\!\!/+M_{ii})}{p^2-M_{ii}^2+{\rm i}\varepsilon}
-2\pi\delta(p^2-M_{ii}^2)(p\!\!\!/+M_{ii})\left[
\vartheta(p_0)f_{N{ii}}(\mathbf{p})
+\vartheta(-p_0)f_{N{ii}}(-\mathbf{p})
\right]\,,\\
{\rm i}S_{N{ii}}^{\bar T}(p)
&=
-\frac{{\rm i}(p\!\!\!/+M_{ii})}{p^2-M_{ii}^2-{\rm i}\varepsilon}
-2\pi\delta(p^2-M_{ii}^2)(p\!\!\!/+M_{ii})\left[
\vartheta(p_0)f_{N{ii}}(\mathbf{p})
+\vartheta(-p_0)f_{N{ii}}(-\mathbf{p})
\right]
\,.
\end{align}
\end{subequations}
In general, $f_N$ decomposes into equilibrium
and out-of-equilibrium contributions as $f_N=f_N^{\rm eq}+\delta f_N$,
where $f_N^{\rm eq}$ is a Fermi-Dirac distribution. For the equilibrium
contributions to ${\rm i}S_N$, we could then derive finite-width
corrections, {\it cf.} Eqs.~(\ref{scalar:finite-width}) and
Ref.~\cite{arXiv:1108.3688} for the scalar case. As these corrections
are not relevant for the present analysis, we do not include
them here.
Notice that the propagators~(\ref{prop:N:expl}) observe
the Majorana constraint~(\ref{Majorana:Propagator}).

We aim for approximate solutions to the
the Kadanoff-Baym equations for the singlet neutrinos,
which  are
\begin{align}
\label{KB:singlet}
&\left(
\slashed k +\frac{\rm i}2 \slashed \partial - M_{ik}
\right)S^{<,>}_{N_{kj}}
-{\rm e}^{-{\rm i}\diamond}\{\slashed\Sigma^H_{N_{ik}}\}\{S^{<,>}_{N_{kj}}\}
-{\rm e}^{-{\rm i}\diamond}\{\slashed\Sigma^{<,>}_{N_{ik}}\}\{S^H_{N_{kj}}\}
\\\notag
=&\frac 12
{\rm e}^{-{\rm i}\diamond}\left(
\{\slashed\Sigma^>_{N_{ik}}\}\{S^<_{N_{kj}}\}
-\{\slashed\Sigma^<_{N_{ik}}\}\{S^>_{N_{kj}}\}
\right)\,,
\end{align}
with the self energy
\begin{align}
\label{Sigma:N}
{\rm i}{\Sigma\!\!\!/}^{ab}_{Nij}(k)
=& \,g_w \int\frac{d^4 k^\prime}{(2\pi)^4}\frac{d^4 k^{\prime\prime}}{(2\pi)^4}
(2\pi)^4 \delta^{(4)}(k-k^\prime-k^{\prime\prime})\\
\notag
&\times
\Big\{
Y_i Y_j^* P_{\rm L} {\rm i}S_{\ell}^{ab}(k^\prime) P_{\rm R} 
{\rm i} \Delta_\phi^{ab}(k^{\prime\prime})
+Y_i^* Y_j C \left[P_{\rm L} {\rm i}S_{\ell}^{ba} (-k^\prime)
P_{\rm R}\right]^t C^\dagger
{\rm i} \Delta_\phi^{ba}(-k^{\prime\prime})
\Big\}\,.
\end{align}
Through the factor $g_w=2$,
we take account of the ${\rm SU}(2)_{\rm L}$ multiplicity.
Since Ref.~\cite{Beneke:2010wd} omits a discussion
of how this self-energy can be obtained
from the two-particle-irreducible (2PI) effective action approach,
that is often used in non-equilibrium field theory,
we briefly present this connection in
Appendix~\ref{appendix:selferg} ({\it cf.} also Ref.~\cite{Jan:2011}).
In analogy with the discussion of the scalar case, the hermitian
part of the self-energy $\slashed\Sigma_N^H$ can eventually be absorbed
in a redefinition of $M^2$, that appears below
in the effective equation~(\ref{evol:neutrino})
for the evolution of the distribution function for the $N_i$.
Technically, this can be performed by generalising the methods for
Weyl fermions in Ref.~\cite{Beneke:2010dz} to the case of massive Majorana
fermions.
The contribution that results from $\slashed\Sigma_N^H$
is real symmetric, but temperature dependent. In general, it is
therefore not possible to diagonalise the effective mass in a
temperature-independent way. The temperature-dependent term is given by
Eq.~(\ref{eff:mass:square}) below.

When replacing the definitions~(\ref{GammaD:scalar})
and~(\ref{DeltaMMBar:scalar}) by
\begin{align}
\label{GammaD:neutrino}
\Gamma_{\rm D}(p)\approx \underset{ij}{\max}
\left|{\rm tr}[\slashed p\slashed\Sigma^{\cal A}_{N_{ij}}(p)]/p^0\right|
\end{align}
and
\begin{subequations}
\label{DeltaMMBar:neutrino}
\begin{align}
\Delta M&=|M_{11}-M_{22}|\,,
\\
\bar M&=\frac{M_{11}+M_{22}}{2}\,,
\end{align}
\end{subequations}
we can identify the same basic approximation
strategies for the calculation of
the lepton asymmetry from
mixing Majorana neutrinos with the calculation of the
asymmetry from mixing neutral
scalars, that is discussed in Section~\ref{section:scalar}.
In particular, it is useful to make use of the approximation
$\Gamma_{\rm D}\ll\Delta M$, where the crucial off-diagonal
components of the singlet neutrino Green functions can be
obtained as small perturbations to the diagonal ones, and
of the approximation $\Delta M\ll \bar M$, where the off-diagonal
Green functions can be obtained from the kinetic equations.
Fortunately, since $\bar M\gg\Gamma_{\rm D}$, due to the
smallness of the couplings $Y$, the parametric regions,
where these conditions hold, overlap,
such that phenomenologically reliable predictions are available
throughout the parameter space.

We should now specify the pivotal quantity $\slashed\Sigma^{\cal A}_N$.
Since for the massless lepton, ${\rm i}S_\ell$ is proportional to $\gamma$-matrices, the
same holds true for the self-energy. In particular, we may write
\begin{align}
{\Sigma\!\!\!/}^{\cal A}_{Nij}(k)
=\gamma^\mu g_w(
Y_i^* Y_j \hat\Sigma^{{\cal A}}_{N{\rm L}\mu} P_{\rm L} 
+ Y_i Y_j^* \hat\Sigma^{{\cal A}}_{N{\rm R}\mu}P_{\rm R})\,.
\end{align}
The expressions for $\hat\Sigma^{{\cal A}\mu}_{N{\rm L,R}}$ with
lepton and Higgs distributions in kinetic equilibrium,
{\it i.e.} for Fermi-Dirac and Bose-Einstein distributions with
chemical potentials $\mu_\ell$ and $\mu_\phi$,
are given by Eqs.~(\ref{Sigma:analytic}). When $\Delta M \gg \Gamma_{\rm D}$
is not fulfilled, the off-diagonal correlations of
the right-handed neutrinos can effectively retain lepton number over
an amount of time that can be relevant for Leptogenesis. In order
to accurately treat the washout of the asymmetry,
it is therefore important to account for the chemical potentials
$\mu_\ell$ and $\mu_\phi$ when solving for the Kadanoff-Baym equations
for the singlet neutrino propagator~(\ref{KB:singlet}).
The extreme
case is a setup considered in Ref.~\cite{Blanchet:2009kk}, where
the right-handed neutrinos and their interactions approximately
conserve lepton number.
On the other hand, when
$\Delta M \gg \Gamma_{\rm D}$ holds, it is a good approximation to
neglect the lepton and Higgs chemical potential for the purpose of
calculating $\slashed\Sigma^{\cal A}_N$, such that we can
set
\begin{align}
\label{SigmaAN:simple}
\Sigma^{{\cal A}\mu}_{N}\equiv
\Sigma^{{\cal A}\mu}_{N_{\rm L}}\big|_{\mu_\ell=\mu_\phi=0}
=\Sigma^{{\cal A}\mu}_{N_{\rm R}}\big|_{\mu_\ell=\mu_\phi=0}
\end{align}
and\footnote{
This definition is related to the similar quantity
in Ref.~\cite{Beneke:2010wd} through
$\hat \Sigma_N^{{\cal A}\mu}=\frac{1}{2g_w} \Sigma_N^\mu$.}
\begin{align}
\label{SigmaAN:nochem}
{\Sigma\!\!\!/}^{\cal A}_{Nij}(k)
=\gamma_\mu\hat\Sigma^{{\cal A}\mu}_N g_w(Y_i Y_j^* P_{\rm R} +Y_i^* Y_j P_{\rm L})\,.
\end{align}
Note that $\frac14{\rm tr}[\slashed k\slashed\Sigma^{\cal A}_{Nii}]/k^0$
is the decay rate of $N_i$.


The one-loop lepton self energy is
\begin{align}
\label{Sigma:ell:wf}
{\rm i}{\Sigma\!\!\!/}^{{\rm wf}<,>}_\ell(k)
=\int\frac{d^4k^\prime}{(2\pi)^4}\frac{d^4k^{\prime\prime}}{(2\pi)^4}
(2\pi)^4\delta^4(k-k^\prime-k^{\prime\prime})
Y_i^*Y_j
P_{\rm R}
{\rm i}S^{{\rm wf}<,>}_{Nij}(k^\prime)
P_{\rm L}
{\rm i}\Delta_\phi^{>,<}(-k^{\prime\prime})\,,
\end{align}
which drives the evolution of the lepton asymmetry according to
\begin{align}
\label{source:washout}
\frac{d}{dt}(n_\ell-\bar n_\ell)
=
S+W
&=\int\frac{d^4 k}{(2\pi)^4}{\rm tr}\left[
{\rm i}{\Sigma\!\!\!/}_\ell^{>}(k)P_{\rm L} {\rm i}S_\ell^{<}(k)
-{\rm i}{\Sigma\!\!\!/}_\ell^{<}(k)P_{\rm L} {\rm i}S_\ell^{>}(k)
\right]
\,,
\end{align}
where $n_\ell$ ($\bar n_\ell$) is the number density of
(anti-) leptons.
We decompose this expression into the source term $S$ and the washout
term $W$. The source is  extracted, when substituting equilibrium
propagators for ${\rm i}S_\ell$ and ${\rm i}\Delta_\phi$. This
approximation is valid, provided the lepton charge density is small
compared to the number density. In particular,
we do not need to account for the lepton and Higgs chemical
potentials when calculating the source term.
The KMS relations then imply, that
the source is proportional to the deviation of the singlet neutrino
from equilibrium,
\begin{align}
\label{source}
S&=Y_i^*Y_j
\int\frac{d^4 k}{(2\pi)^4}
\int\frac{d^4 k^\prime}{(2\pi)^4}
\int\frac{d^4 k^{\prime\prime}}{(2\pi)^4}
(2\pi)^4\delta^4(k-k^\prime-k^{\prime\prime})
\\\notag
&\hskip2cm\times
{\rm tr}\left[
P_{\rm R} {\rm i} \delta S_{N_{ij}}(k^\prime) P_{\rm L}
\left(
{\rm i}\Delta^<_\phi(-k^{\prime\prime})
{\rm i}S_\ell^{<}(k)
-{\rm i}\Delta^>_\phi(-k^{\prime\prime})
{\rm i}S_\ell^{>}(k)
\right)
\right]
\\\notag
&=
-Y_i^*Y_j\int\frac{d^4 k^\prime}{(2\pi)^4}
{\rm tr}\left[
P_{\rm R} {\rm i} \delta S_{N_{ij}}(k^\prime)\, 2 P_{\rm L}
\hat{\slashed{\Sigma}}_N^{\cal A}(k^\prime)
\right]
\,,
\end{align}
with equilibrium distributions and vanishing
chemical potentials for $\ell$ and $\phi$,
such that $\hat{{\Sigma}}_N^{\cal A}$ is given by
Eq.~(\ref{SigmaAN:nochem}).
A complete
network of equations that determine the evolution of the
lepton charge in the expanding Universe is presented in
Section~\ref{section:effective:theory}.

\subsection{Helicity Block-Diagonal Decomposition of the Singlet Neutrino Propagator}

In order to explicitly demonstrate the consistency of the constraint
and the kinetic equations when $\Delta M\gg \Gamma_{\rm D}$
and in order to pave the ground for the calculation of
the asymmetry when $\Delta M\ll \bar M$, we decompose the
fermionic Green functions into various Dirac matrix components.
In particular, for a spatially homogeneous and isotropic
problem, it is useful to
notice that the helicity operator $\hat h=\hat k^i\gamma^0\gamma^i\gamma^5$
commutes with the various Dirac matrices that appear in the
Kadanoff-Baym equations~(\ref{KB:singlet}), such that helicity is
a good quantum number. This suggests the
decomposition~\cite{Prokopec:2003pj,hep-ph/0105295,hep-th/0211219}
\begin{align}
\label{helicity:decomposition}
{\rm i}\delta S_N=\sum\limits_{h=\pm}{\rm i}\delta S_{Nh}\,,
\qquad
-{\rm i}\gamma^0 \delta S_{Nh}=\frac14(\mathbbm 1+h \hat k^i \sigma^i)
\otimes \rho^a g_{ah}
\end{align}
where $h=\pm$ denotes helicity and $\sigma^i$, $\rho^i$ are Pauli
matrices.
Note that the functions $g_{ah}$ are understood to
be hermitian matrices in flavour space. It may be useful to compare
with Eq.~(\ref{Bloch:explicit}), in order to acquire a quick
understanding
of the functions $g_{ah}$.

Multiplication of the Kadanoff-Baym equations~(\ref{KB:singlet})
by $\{\mathbbm 1,-h\gamma^i\gamma^5,-{\rm i} h \gamma^i,-\gamma^5\}$, taking
the Dirac trace and truncating
the diamond operators yields the equations
\begin{align}
\label{KB:matrix}
\left[
{\cal F}_{h}-{\bf Y}^*{\bf Y}^t{\cal C}_{h\,{\rm L}}-{\bf Y}{\bf Y}^\dagger{\cal C}_{h\,{\rm R}}
\right]
\left(
\begin{array}{c}
g_{0h}\\
g_{1h}\\
g_{2h}\\
g_{3h}
\end{array}
\right)
=0\,,
\end{align}
where ${\bf Y}^t=(Y_1,Y_2)$,
\begin{align}
{\cal F}_{h}=
\left(
\begin{array}{cccc}
-{\rm i k^0} +\frac12\partial_t & {\rm i} M & 0 & {\rm i} h |\mathbf k|\\
{\rm i} M&-{\rm i k^0}+\frac12\partial_t & h |\mathbf k| & 0\\
0 & - h |\mathbf k| &-{\rm i k^0} +\frac12\partial_t& M\\
{\rm i} h |\mathbf k| & 0 & -M &-{\rm i k^0}+\frac12\partial_t\\
\end{array}
\right)\,,
\end{align}
and
\begin{align}
&{\cal C}_{h\,{\rm L},{\rm R}}=
\frac{g_w}2\times
\\\nonumber
&\!\!\!\!\!
\left(
\begin{array}{cccc}
-\hat\Sigma_{N{\rm L,R}}^{{\cal A}0}\mp h \hat k^i\hat\Sigma_{N{\rm L,R}}^{{\cal A}i}	&	0	&	0	&\pm\hat\Sigma_{N{\rm L,R}}^{{\cal A}0} +h\hat k^i\hat\Sigma_{N{\rm L,R}}^{{\cal A}i}	\\
			0	&\;-\hat\Sigma_{N{\rm L,R}}^{{\cal A}0}\mp h\hat k^i\hat\Sigma_{N{\rm L,R}}^{{\cal A}i}\;&\;\mp{\rm i}\hat\Sigma_{N{\rm L,R}}^{{\cal A}0}-{\rm i}h\hat k^i\Sigma_{N{\rm L,R}}^{{\cal A}i}\; & 0\\
0 &\;\pm{\rm i}\hat\Sigma_{N{\rm L,R}}^{{\cal A}0}+{\rm i}h\hat k^i\hat\Sigma_{\rm L,R}^{{\cal A}i}\; &\; -\hat\Sigma_{N{\rm L,R}}^{{\cal A}0}\mp h \hat k^i\hat\Sigma_{N{\rm L,R}}^{{\cal A}i}\; & 0\\
\pm\hat\Sigma_{N{\rm L,R}}^{{\cal A}0} +h\hat k^i\hat\Sigma_{N{\rm L,R}}^{{\cal A}i} & 0 & 0 & -\hat\Sigma_{N{\rm L,R}}^{{\cal A}0} \mp h \hat k^i\hat\Sigma_{N{\rm L,R}}^{{\cal A}i}
\end{array}
\right)
\,.
\end{align}

The kinetic equations correspond to the hermitian part of
Eqs.~(\ref{KB:matrix}), which is
\begin{subequations}
\label{kin:eqs}
\begin{align}
\label{kin:eq:f0}
\dot g_{0h} +{\rm i}[M,g_{1h}]&=
\sum_\pm
-\frac{g_w}2
\left(
\hat\Sigma_{N{\rm L,R}}^{{\cal A}0}\pm h\hat k^i\hat\Sigma_{N{\rm L,R}}^{{\cal A}i}
\right)
\{
\Upsilon_{{\rm L},{\rm R}},
g_{0h}\mp g_{3h}
\}
\,,
\\
\dot g_{1h} +2h|\mathbf k|g_{2h}+{\rm i}[M,g_{0h}]&=
\sum_\pm
\frac{g_w}2
\left(
\hat\Sigma_{N{\rm L,R}}^{{\cal A}0}\pm h\hat k^i\hat\Sigma_{N{\rm L,R}}^{{\cal A}i}
\right)
\left(
-\{
\Upsilon_{{\rm L},{\rm R}},
g_{1h}
\}
\mp{\rm i}
[
\Upsilon_{{\rm L},{\rm R}},
g_{2h}
]
\right)
\,,
\\
\label{kin:eq:f2}
\dot g_{2h} -2h|\mathbf k|g_{1h}+\{M,g_{3h}\}&=
\sum_\pm
\frac{g_w}2
\left(
\hat\Sigma_{N{\rm L,R}}^{{\cal A}0}\pm h\hat k^i\hat\Sigma_{N{\rm L,R}}^{{\cal A}i}
\right)
\left(
\pm{\rm i}
[
\Upsilon_{{\rm L},{\rm R}},
g_{1h}
]
-\{
\Upsilon_{{\rm L},{\rm R}},
g_{2h}
\}
\right)
\,,\\
\dot g_{3h} -\{M,g_{2h}\}&=
\sum_\pm
-\frac{g_w}2
\left(
\hat\Sigma_{N{\rm L,R}}^{{\cal A}0}\pm h\hat k^i\hat\Sigma_{N{\rm L,R}}^{{\cal A}i}
\right)
\{
\Upsilon_{{\rm L},{\rm R}},
g_{3h}\mp g_{0h}
\}
\,,
\end{align}
\end{subequations}
where we define the shorthand notations
\begin{align}
\Upsilon_{\rm L}=\mathbf Y^* \mathbf Y^t\,,\qquad
\Upsilon_{\rm R}=\mathbf Y \mathbf Y^\dagger\,.
\end{align}
The
constraint equations are the anti-hermitian part of Eqs.~(\ref{KB:matrix}) and read
\begin{subequations}
\label{constr:eqs}
\begin{align}
\label{constr:eq:0}
-2{\rm i} k^0 g_{0h} +2{\rm i}h|\mathbf k|g_{3h}+{\rm i}\{M,g_{1h}\}
&=
\sum_\pm
-\frac{g_w}2
\left(
\hat\Sigma_{N{\rm L,R}}^{{\cal A}0}\pm h\hat k^i\hat\Sigma_{N{\rm L,R}}^{{\cal A}i}
\right)
[
\Upsilon_{{\rm L},{\rm R}},
g_{0h}\mp g_{3h}
]
\,,
\\
\label{constr:eq:1}
-2{\rm i} k^0 g_{1h} +{\rm i}\{M,g_{0h}\}&=
\sum_\pm
\frac{g_w}2
\left(
\hat\Sigma_{N{\rm L,R}}^{{\cal A}0}\pm h\hat k^i\hat\Sigma_{N{\rm L,R}}^{{\cal A}i}
\right)
\\\notag
&\times
\left(
-[
\Upsilon_{{\rm L},{\rm R}},
g_{1h}
]
\mp{\rm i}
\{
\Upsilon_{{\rm L},{\rm R}},
g_{2h}
\}
\right)
\,,
\\
\label{constr:eq:2}
-2{\rm i} k^0 g_{2h} +[M,g_{3h}]&=
\sum_\pm
\frac{g_w}2
\left(
\hat\Sigma_{N{\rm L,R}}^{{\cal A}0}\pm h\hat k^i\hat\Sigma_{N{\rm L,R}}^{{\cal A}i}
\right)
\\\notag
&\times
\left(
\pm{\rm i}
\{
\Upsilon_{{\rm L},{\rm R}},
g_{1h}
\}
-[
\Upsilon_{{\rm L},{\rm R}},
g_{2h}
]
\right)
\,,\\
\label{constr:eq:3}
-2{\rm i} k^0 g_{3h}+2{\rm i}h|\mathbf k|g_{0h} -[M,g_{2h}]&=
\sum_\pm
-
\frac{g_w}2
\left(
\hat\Sigma_{N{\rm L,R}}^{{\cal A}0}\pm h\hat k^i\hat\Sigma_{N{\rm L,R}}^{{\cal A}i}
\right)
[
\Upsilon_{{\rm L},{\rm R}},
g_{3h}\mp g_{0h}
]
\,.
\end{align}
\end{subequations}

\subsection{Derivation of the Asymmetry for $\Delta M\gg\Gamma_{\rm D}$}

In the regime $\Delta M\gg\Gamma_{\rm D}$, we calculate the
off-diagonal components of ${\rm i}\delta S_{Nij}$
as small perturbations for a given diagonal out-of-equilibrium
distribution. We verify that constraint and kinetic equations
give consistent answers.
This is in analogy with the procedure that
is presented in Section~\ref{section:scalar:nondeg} for the scalar case,
and the
error is controlled by the expansion parameter
$\Gamma_{\rm D}/\Delta M$.

We first derive the source of the lepton asymmetry from the
constraint equations~(\ref{constr:eqs}).
Assuming a mass-diagonal basis and neglecting the collision term,
we obtain from the diagonal components of Eqs.~(\ref{constr:eqs})
the relations
\begin{align}
\label{sol:diag}
g_{1hii}(k)=\frac{M_{ii}}{k^0}g_{0hii}(k)\,,\qquad
g_{2hii}(k)=0\,,\qquad
g_{3hii}(k)=\frac{h|\mathbf k|}{k^0}g_{0hii}(k)
\,.
\end{align}
For these diagonal components, we assume
helicity-independence, {\it i.e.} $g_{0+ii}=g_{0-ii}$.
This follows when
assuming that there is no initial helicity asymmetry within the singlet neutrinos 
and no initial charge asymmetry within the lepton and Higgs sector (and
hence, no $CP$-asymmetry). Because the tree-level decays and inverse decays
then produce no helicity asymmetry, also the non-equilibrium distributions
for the $N_i$ are helicity symmetric at leading order. A helicity asymmetry in $N_i$ is
only incurred through $CP$-violating loop effects and the back-reaction of the
produced lepton asymmetry, which results in sub-leading contributions
to the non-equilibrium distributions for the $N_i$. Note that symmetric initial conditons
are naturally established within the parametric range of strong washout.
Through Eqs.~(\ref{prop:N:expl}),
we may relate $g_{0hii}$ to the distribution functions as
\begin{align}
{\rm sign}(k^0)2\sqrt{\mathbf k^2+M_{ii}^2}\delta f_{N_{ii}}(\mathbf k)2\pi\delta(k^2-M_{ii}^2)
=g_{0hii}(k)\,.
\end{align}
From the expression for the source term~(\ref{source}),
we see that we need to determine
the off-diagonal components (that is
$i\not=j$) of
$P_{\rm R}{\rm i}\delta S_{N ij} P_{\rm L}$.
Therefore,
we employ the diagonal solutions to derive the off-diagonal contributions
as perturbations of order $\Gamma_{\rm D}/\Delta M$.
Because $\Delta M\gg \Gamma_{\rm D}$, lepton number is rapidly violated
once the inverse decay into a right-handed neutrino has occurred.
We can therefore neglect the lepton and Higgs chemical potentials
when calculating the singlet propagator and use the approximation~(\ref{SigmaAN:simple}).
The constraint equations
for the off-diagonal components ($i\not=j$) are then given by
\begin{subequations}
\label{constr:eqs12}
\begin{align}
\label{constr:eq12:0}
-2{\rm i} k^0 g^{{\rm L},{\rm R}}_{0hij} +2{\rm i}h|\mathbf k|g^{{\rm L},{\rm R}}_{3hij}+{\rm i}(M_{ii}+M_{jj})g^{{\rm L},{\rm R}}_{1hij}
&=
\frac{1}2
\left(
\Sigma_{Nij}^{{\cal A}0}\pm h\hat k^i\Sigma_{Nij}^{{\cal A}i}
\right)
(
g_{0hii}\mp g_{3hii}
)
\,,
\\
\label{constr:eq12:1}
-2{\rm i} k^0 g^{{\rm L},{\rm R}}_{1hij}
+{\rm i}(M_{ii}+M_{jj})g^{{\rm L},{\rm R}}_{0hij}&=
\frac{1}2
\left(
\Sigma_{Nij}^{{\cal A}0}\pm h\hat k^i\Sigma_{Nij}^{{\cal A}i}
\right)
(
g_{1hii}
\mp{\rm i}
g_{2hii}
)
\,,
\\
\label{constr:eq12:2}
-2{\rm i} k^0 g^{{\rm L},{\rm R}}_{2hij} +(M_{ii}-M_{jj})g^{{\rm L},{\rm R}}_{3hij}&=
\pm
\frac{1}2
{\rm i}
\left(
\Sigma_{Nij}^{{\cal A}0}\pm h\hat k^i \Sigma_{Nij}^{{\cal A}i}
\right)
(
g_{1hii}
\mp{\rm i}
g_{2hii}
)
\,,\\
\label{constr:eq12:3}
-2{\rm i} k^0 g^{{\rm L},{\rm R}}_{3hij}+2{\rm i}h|\mathbf k|g^{{\rm L},{\rm R}}_{0hij} -(M_{ii}-M_{jj})g^{{\rm L},{\rm R}}_{2hij}&=
\mp
\frac{1}2
\left(
\Sigma_{Nij}^{{\cal A}0}\pm h\hat k^i \Sigma_{Nij}^{{\cal A}i}
\right)
(
g_{0hii}\mp g_{3hii}
)
\,,
\end{align}
\end{subequations}
where
$g^{{\rm L}}_{0,3hij}=Y_i^*Y_j \hat g^{{\rm L}}_{0,3hij}$,
$g^{{\rm R}}_{0,3hij}=Y_j^*Y_i \hat g^{{\rm R}}_{0,3hij}$
and $g_{0,3hij}=g^{{\rm L}}_{0,3hij}+g^{{\rm R}}_{0,3hij}$.
Because of hermiticity,
$g^{{\rm L},{\rm R}}_{0,3hji}=(g^{{\rm L},{\rm R}}_{0,3hij})^*$.
It is assumed here, that only $g_{hii}\not=0$, while
$g_{hjj}=0$. The general solution can be constructed by superposition.

Given the diagonal solutions~(\ref{sol:diag}), we can now
straightforwardly determine the $g_{ahij}$ for $i\not=j$
as solutions to the system of linear equations~(\ref{constr:eqs12}).
In particular, the flavour off-diagonal vector- and pseudovector-densities are
\begin{subequations}
\label{f:nondeg:offdiag}
\begin{align}
\hat g_{0h12}^{{\rm L},{\rm R}}&=
\frac{M_{11}(M_{11}+M_{22})\mp2h|\mathbf k|k^0+2\mathbf k^2}
{k^0 (M_{11}^2-M_{22}^2)}
{\rm i}\frac{g_w}{2}\left(
\hat\Sigma^{{\cal A}0}_N
\pm h \hat k^i\hat\Sigma^{{\cal A}i}_N
\right)
g_{0h11}\,,
\\
\hat g^{{\rm L},{\rm R}}_{3h12}&=
\frac{M_{11}(M_{11}-M_{22})\mp2h|\mathbf k|k^0+2\mathbf k^2}
{k^0 (M_{11}^2-M_{22}^2)}
{\rm i}\frac{g_w}{2}\left(
\mp\hat\Sigma^{{\cal A}0}_N
-h \hat k^i\hat\Sigma^{{\cal A}i}_N
\right)
g_{0h11}\,.
\end{align}
\end{subequations}
We have now assumed that only $N_1$ deviates from equilibrium.
As stated above,
the general case, where $N_2$ deviates as well, can
simply be obtained
by superposition.

Note that these sub-leading off-diagonal solutions do not obey
the leading-order symmetry property, according to which
$g_{0h}$ is helicity-even whereas $g_{3h}$ is helicity-odd.
For the $CP$-violating source, above results combine to a compact form
that observes the symmetries of the problem:
\begin{align}
\label{deltaSij}
Y_i^*Y_jP_{\rm R} {\rm i} \delta S_{N_{ij}}(k) P_{\rm L}
&=-Y_i^*Y_j\frac 14\sum_{h=\pm}
\left[
P_{\rm R} \gamma^0 P_{\rm L}-hP_{\rm R} \hat k^i \gamma^i P_{\rm L}
\right]
(g_{0hij}+g_{3hij})
\\\notag
&=-({Y_1^*}^2Y_2^2-Y_1^2 {Y_2^*}^2)\frac{\rm i}{k^0}
\frac{M_{11} M_{22}}{M_{11}^2-M_{22}^2}
P_{\rm R}\gamma^\mu P_{\rm L}
\frac{g_w}2
\hat\Sigma^{\cal A}_{N\mu}
g_{0h11}
\\\notag
&=-({Y_1^*}^2Y_2^2-Y_1^2 {Y_2^*}^2) {\rm i}
\frac{M_{11} M_{22}}{M_{11}^2-M_{22}^2}
\frac{g_w}2
P_{\rm R}\hat{\slashed\Sigma}^{\cal A}_{N}P_{\rm L}
2\delta f_{N11}(\mathbf k) 2\pi\delta(k^2-M_1^2)\,.
\end{align}
Since $N_2$ is assumed to be in equilibrium,
$\delta f_{N22}=0$. Notice that only contributions
from $\hat g^{\rm L}_{0h11}$ enter here, while those from
$\hat g^{\rm R}_{0h11}$ cancel since they are multiplied by the conjugate
Yukawa couplings.
Substituting Eq.~(\ref{deltaSij}) into Eq.~(\ref{source}), we obtain the
known result for the $CP$-violating source of the lepton
asymmetry~\cite{Beneke:2010wd},
\begin{align}
\label{source:well-known}
S=\int \frac{d^3k}{(2\pi)^3 2\sqrt{\mathbf{k}^2+M_{11}^2}}
8{\rm i}[{Y_1^*}^2Y_2^2-Y_1^2 {Y_2^*}^2]\frac{M_{11} M_{22}}{M_{11}^2-M_{22}^2}
g_w
\hat\Sigma^{\cal A}_{N\mu}\hat\Sigma_N^{{\cal A}\mu}
\delta f_{N_{11}}(\mathbf k)
\,,
\end{align}
which includes the finite-density corrections. We refer to
this expression as the perturbative result for the source term,
as it can be calculated as a perturbation to the diagonal singlet
neutrino propagator or, alternatively, from a loop expansion
of Feynman diagrams as in Ref.~\cite{Beneke:2010wd}.

While above result is obtained from the constraint
equations~(\ref{constr:eqs}),
we now check the
consistency with the kinetic equations~(\ref{kin:eqs}) when
using the same approximations:
When $\Delta M\gg\Gamma_{\rm D}$, the diagonal $(11)$-densities track the deviation
of $N_1$ from equilibrium, which provides the thermodynamical
breakdown of time-reversal invariance, which is necessary for
baryogenesis. Compared to these diagonal components, the off-diagonal
ones are suppressed by a factor of order $\Gamma_{\rm D}/\Delta M$.
In the absence of flavour oscillations,
the time-derivatives of the off-diagonals are additionally
suppressed by the same parameter, such that we may neglect them.
With these approximations,
the resulting set of equations obtained from the (12)-components
of Eqs.~(\ref{kin:eqs}) is
\begin{subequations}
\begin{align}
{\rm i}(M_{11}-M_{22}) g^{{\rm L},{\rm R}}_{1h12}&=
-\frac{1}2
(\Sigma^{{\cal A}0}_{N12}\pm h\hat k^i \Sigma^{{\cal A}i}_{N12})
(g_{0h11}\mp g_{3h11})\,,
\\
2h|\mathbf k|g^{{\rm L},{\rm R}}_{2h12}+{\rm i}(M_{11}-M_{22})g^{{\rm L},{\rm R}}_{0h12}
&=
-\frac{1}2
(\Sigma^{{\cal A}0}_{N12}\pm h\hat k^i \Sigma^{{\cal A}i}_{N12})
(g_{1h11}\mp {\rm i}g_{2h11})\,,
\\
-2h|\mathbf k|g^{{\rm L},{\rm R}}_{1h12}+(M_{11}+M_{22})g^{{\rm L},{\rm R}}_{3h12}
&=\mp\frac{1}2
{\rm i}(\Sigma^{{\cal A}0}_{N12}\pm h\hat k^i \Sigma^{{\cal A}i}_{N12})
(g_{1h11}\mp {\rm i}g_{2h11})\,,
\\
-(M_{11}+M_{22})g^{{\rm L},{\rm R}}_{2h12}
&=\pm\frac{1}2
(\Sigma^{{\cal A}0}_{N12}\pm h\hat k^i \Sigma^{{\cal A}i}_{N12})
(g_{0h11}\mp g_{3h11})\,.
\end{align}
\end{subequations}
Indeed, from these equations, we can
straightforwardly reproduce the
result~(\ref{f:nondeg:offdiag}).

\subsection{Asymmetry for $\Delta M \ll \bar M$}

In order
to obtain a solution for the $CP$-violating source also
in the regime where $\Delta M$ is of the same order as $\Gamma_{\rm D}$
or smaller,
we could aim
for a full solution of the kinetic equations~(\ref{kin:eqs}).
Within an intuitive picture of flavour oscillations, the meaning
of Eqs.~(\ref{kin:eqs}) is yet somewhat obscure because of the
mixing of the $g_{ah}$, which is a result of the spinor structure.
A diagonalisation of this spinor structure is not possible in general,
because the dispersive contributions from $M$ and the absorptive ones
from $\slashed\Sigma_N^{\cal A}$ cannot be diagonalised simultaneously.

In the close-to-degenerate regime, where flavour oscillations are
important,
crucial simplifications can however be achieved. We do not need
to specify the locations of the quasi-particle poles of the
diagonal and off-diagonal components of the propagators exactly,
but only make use of the information that these are located at
$p^2=\bar M^2$, up to a relative error of order $\Delta M/\bar M$.
We can then integrate the constraint
and the kinetic equations~(\ref{kin:eqs})
over an interval ${\cal I}$ of length of order $\Delta M$
over the region of the quasi-particle poles. As a result, we
obtain again the kinetic equations~(\ref{kin:eqs}) and
the constraint equations~(\ref{constr:eqs}) with
the replacement $g_{ahij}\to \delta f_{ahij}$, where
\begin{align}
\label{f:fermi}
\int\limits_{{\cal I}_\pm}\frac{d p^0}{2\pi}
{\rm sign}(p^0) g_{ahij}(p)
=\left\{
\begin{array}{l}
\delta f_{ahij}(\mathbf p)\quad\textnormal{for}\quad+\\
\delta \bar f_{ahij}(\mathbf p)\quad\textnormal{for}\quad-\\
\end{array}
\right.
\,,
\end{align}
and ${\cal I}_\pm$ is again an interval around the positive or
negative energy quasi-particle pole, respectively.
The error results from the averaging over values of $p^0$ in the constraint
equations and can be estimated to be of relative order $\Delta M/\bar M$.
In order to keep the notation compact when we have to
distinguish the cases of positive and negative frequencies,
we adapt the pragmatic definition~(\ref{def:fp}) to the
fermionic case,
\begin{align}
\label{def:fp:fermi}
\delta f_{ahij}(p)=\left\{
\begin{array}{l}
\delta f_{ahij}(\mathbf p)\;\;{\rm for}\;\;p^0>0\\
\delta \bar f_{ahij}(\mathbf p)\;\;{\rm for}\;\;p^0<0
\end{array}
\right.
\,.
\end{align}

When
$\Gamma_{\rm D}\ll \bar M$ and $\Delta M\ll \bar M$, it follows from
Eq.~(\ref{kin:eq:f2}) or from Eqs.~(\ref{constr:eqs})
that
\begin{align}
\label{constrf1f3}
\delta f_{1hij}(k)=\delta f_{3hij}(k)\frac{M_{ii}+M_{jj}}{2h|\mathbf k|}
\end{align}
and from Eq.~(\ref{constr:eq:1}) that
\begin{align}
\label{constrf1f0}
\delta f_{1hij}(k)=\delta f_{0hij}(k)\frac{M_{ii}+M_{jj}}{2k^0}\,.
\end{align}

Using these approximate constraints, Eq.~(\ref{kin:eq:f0})
becomes
\begin{align}
\label{flavour:dyn:fermi}
\delta \dot f_{0h}+\frac{1}{2k^0}{\rm i}[M^2,\delta f_{0h}]
=&
-
\sum_\pm
\frac{g_w}2
\left(\hat\Sigma^{{\cal A}0}_{N{\rm L,R}}\pm h\hat k^i\hat\Sigma^{{\cal A}i}_{N{\rm L,R}}
\right)
\left(1\mp \frac{h|\mathbf k|}{k^0}\right)
\left\{
\Upsilon_{{\rm L},{\rm R}},
\delta f_{0h}
\right\}\\
=&
- g_w \bigg\{{\rm Re}[\mathbf Y^* \mathbf Y^t]\frac{k\cdot\hat\Sigma^{\cal A}_{N}}{k^0} - {\rm i}h{\rm Im}[\mathbf Y^* \mathbf Y^t]\frac{\tilde k\cdot\hat\Sigma^{\cal A}_{N}}{k^0},\delta f_{0h}\bigg\}
\nonumber\,,
\end{align}
where $\tilde k \equiv (|\mathbf k|, k^0 \hat{k})$ and in the last row we have
 neglected the lepton and Higgs chemical potentials according to
 Eq.~(\ref{SigmaAN:simple}). As discussed above, the latter approximation is not in
 general justified when $\Delta M \lesssim \Gamma_{\rm D}$.
The off-diagonal components of the out-of-equilibrium
Wightman function for the singlet neutrinos can then be
approximated as
\begin{align}
g_{ahij}(p)=2\pi \delta(p^2-\bar M^2) 2p^0
\delta f_{ahij}(p)\,,
\end{align}
which is accurate up to a relative error of order
$\Delta M/\bar M$. In order to calculate the
$CP$-asymmetric source, this should then be substituted in
Eq.~(\ref{helicity:decomposition}) and eventually in
Eq.~(\ref{source}).

We can easily perform the consistency check that Eq.~(\ref{flavour:dyn:fermi})
reproduces above results when
$\Gamma_{\rm D}\ll \bar M$ and $\Delta M\ll \bar M$ but
$\Delta M\gg\Gamma_{\rm D}$. Neglecting the lepton and Higgs chemical potentials
according to Eq.~(\ref{SigmaAN:simple}), the off-diagonal terms that arise from
the out-of-equilibrium singlet neutrino $N_1$ are
\begin{subequations}
\begin{align}
\label{offdiag:degen:noosc}
\delta f_{0h12}&=2{\rm i}\bigg[Y_1^*Y_2 \frac{k^0- h|\mathbf k|}{M_{11}^2-M_{22}^2}
\frac{g_w}2
\left(
\hat\Sigma^{{\cal A}0}_{12}+ h\hat k^i\hat\Sigma^{{\cal A}i}_{12}
\right)
\\\notag
&
\hskip.6cm
+
Y_1Y_2^* \frac{k^0+ h|\mathbf k|}{M_{11}^2-M_{22}^2}
\frac{g_w}2
\left(
\hat\Sigma^{{\cal A}0}_{N}- h\hat k^i\hat\Sigma^{{\cal A}i}_{N}
\right)
\bigg]
\delta f_{0h11}\,,
\\
\label{offdiag:degen:noosc3}
\delta f_{3h12}&=
2{\rm i}
\bigg[
Y_1^*Y_2\frac{\mathbf k^2 - h|\mathbf k|k^0}{k^0(M_{11}^2-M_{22}^2)}
\frac{g_w}2
\left(
-\hat\Sigma^{{\cal A}0}_{N}- h\hat k^i\hat\Sigma^{{\cal A}i}_{N}
\right)
\\\notag
&
\hskip.6cm
+
Y_1Y_2^*\frac{\mathbf k^2 + h|\mathbf k|k^0}{k^0(M_{11}^2-M_{22}^2)}
\frac{g_w}2
\left(
\hat\Sigma^{{\cal A}0}_{N}- h\hat k^i\hat\Sigma^{{\cal A}i}_{N}
\right)
\bigg]
\delta f_{0h11}\,,
\end{align}
\end{subequations}
where we have neglected the time-derivatives on the left-hand side
of Eq.~(\ref{flavour:dyn:fermi}). Therefore, these equations
cannot take into account possible flavour oscillations
for the $N_i$, which should however be irrelevant for
the lepton asymmetry when $\Delta M\gg\Gamma_{\rm D}$, as
we explain in Section~\ref{section:illustrative}.
Note that these expressions can as well be obtained from
Eqs.~(\ref{f:nondeg:offdiag}) when taking $M_{11}\to M_{22}$
and using the definition~(\ref{f:fermi}). As a further
consistency check,
we also note that
in the regime where $\Delta M\gg\Gamma_{\rm D}$
but $\Delta M\ll \bar M$,
we can exchange $M_{11}\leftrightarrow M_{22}$ within Eqs.~(\ref{sol:diag})
and $M_{11}+M_{22}\leftrightarrow 2 M_{11,22}$ in Eqs.~(\ref{kin:eqs}).
Within the result~(\ref{f:nondeg:offdiag}) for
$g_{0h12}$ and $g_{3h12}$, this indeed
only incurs a relative error of order
$\Delta M/\bar M$, which is a small correction.


Now, when we relax the requirement that $\Delta M\gg\Gamma_{\rm D}$,
Eqs.~(\ref{flavour:dyn:fermi}) can be straightforwardly solved numerically.
Besides, it is also instructive to obtain analytical solutions
in a situation without expansion of the Universe. These
can be obtained when substituting
\begin{align}
\label{Omega:Gamma:fermi}
\Omega=\frac{M^2}{k^0}\qquad
\textnormal{and}
\qquad
\Gamma=&2\sum_\pm
\frac{g_w}2
\Upsilon_{{\rm L},{\rm R}}
\left(
\hat\Sigma^{{\cal A}0}_{N{\rm L,R}}\pm h \hat k^i\hat\Sigma^{{\cal A}i}_{N{\rm L,R}}
\right)
\left(1\mp\frac{h|\mathbf k|}{k^0}\right)
\\ =& 2 g_w \bigg({\rm Re}[\mathbf Y^* \mathbf Y^t]\frac{k\cdot\hat\Sigma^{\cal A}_{N}}{k^0} - {\rm i}h{\rm Im}[\mathbf Y^* \mathbf Y^t]\frac{\tilde k\cdot\hat\Sigma^{\cal A}_{N}}{k^0}\bigg)
\nonumber
\end{align}
in Eq.~(\ref{flavour:dynamic:general}).
In the last row of Eq.~(\ref{Omega:Gamma:fermi}) we have again
 neglected the lepton and Higgs chemical potentials according to
 Eq.~(\ref{SigmaAN:simple}).
Before we discuss some features of these solutions in
Section~\ref{section:illustrative}, we note an additional
consistency check.
When neglecting
oscillatory contributions, as it is appropriate for
$\Gamma_{\rm D}\ll\Delta M$, we can immediately
recover the perturbative solution~(\ref{offdiag:degen:noosc})
from Eq.~(\ref{offdiag:analytical}).
Eq.~(\ref{offdiag:degen:noosc3})
is then obtained through $g_{3h}=(h|\mathbf k|/k^0)g_{0h}$, which
also holds for the flavour off-diagonals in the close-to-degenerate regime,
as we have demonstrated above in
Eqs.~(\ref{constrf1f3},\ref{constrf1f0}).
When $\Delta M \gg \Gamma_{\rm D}$,
we therefore consistently reproduce the result for the lepton number
violating contribution to the off-diagonal singlet-neutrino
Wightman function~(\ref{deltaSij}) and eventually the well-known result for
the $CP$-violating source~(\ref{source:well-known}).

We have therefore
established that for $\Delta M\gg\Gamma_{\rm D}$, the kinetic
equations~(\ref{flavour:dynamic:general}) accurately reproduce
the classic result for the resonantly enhanced lepton
asymmetry~\cite{Flanz:1996fb,Covi:1996wh,Pilaftsis:1997dr,Pilaftsis:1997jf,Pilaftsis:2003gt,Pilaftsis:2005rv},
that arises as the zero-temperature limit of Eq.~(\ref{source:well-known}). In order to achieve this agreement, the helicity structure
exhibited in Eqs.~(\ref{flavour:dyn:fermi})
and~(\ref{Omega:Gamma:fermi}) is crucial. For the lepton
asymmetry from decaying Majorana neutrinos, to our knowledge, this is
the first time an explicit connection between results
from the $S$-matrix  approach
and from kinetic evolution equations is established.

\section{Illustrative Features of
Analytical Solutions when Ignoring
Washout and the Expanding Background}
\label{section:illustrative}

We specialise here to the regime where the approximation
$\Delta M\ll M$ is valid and first consider the
scalar model. Therefore,
we use the form~(\ref{OS:scalar:resonant})
for the off-diagonal Wightman functions of $\chi$. The
source term for the asymmetry~(\ref{asymmetry:compact}) can then be
expressed as
\begin{align}
S_\varphi=-g_i g_j^*
\int\frac{d^3 q}{(2\pi)^3 2\sqrt{\mathbf q^2 +\bar M^2}}
\sum\limits_{q^0=\pm \omega(\mathbf q)}
\hat\Pi_\chi^{\cal A}(q)\delta f_{\chi_{ij}}(q)\,.
\end{align}
When adding the positive and negative energy contributions, it is useful to
notice the transformation properties of the quantities in
Eqs.~(\ref{diagonalisation:parameters})
\begin{subequations}
\begin{align}
\Delta&\underset{q^0\to -q^0}\longrightarrow-\Delta^*\,,\\
D&\underset{q^0\to -q^0}\longrightarrow-D^*\,,\\
\Xi_{{\rm D}ii}&\underset{q^0\to -q^0}\longrightarrow-\Xi_{{\rm D}ii}^*\,.
\end{align}
\end{subequations}
As a consequence, $\delta f_{\chi_{ij}}(k^0,\mathbf k)=\delta f_{\chi_{ji}}(-k^0,\mathbf k)$,
a relation that can also be derived from the neutrality
condition~(\ref{neutrality:scalar}) imposed
on ${\rm i}\Delta_\chi$.
Use of the result~(\ref{densities:analytical}) then yields
\begin{align}
S_\varphi&=
-{\rm i}(g_1^2{g_2^*}^2-{g_1^*}^2g_2^2)
\int\frac{d^3 q}{(2\pi)^3 2\sqrt{{\mathbf q}^2+\bar M^2}}
\delta f_0(\mathbf q)
\frac{{\hat{\Pi}_\chi^{{\cal A}^2}}}{\left|(\Delta+D)^2-|\Pi^{\cal A}_{\chi_{12}}/\omega(\mathbf q)|^2\right|^2}
\\\notag
&\times
\left[
(\Delta^*+D^*)
\left(
{\rm e}^{\frac{\rm i}2\Xi^{\rm c}_{{\rm D}11}t}
-{\rm e}^{\frac{\rm i}2\Xi^{\rm c}_{{\rm D}22}t}
\right)
\left(
(\Delta+D)^2
{\rm e}^{-\frac{\rm i}2\Xi_{{\rm D}11}t}
-|\Pi^{\cal A}_{\chi_{12}}/\omega(\mathbf q)|^2
{\rm e}^{-\frac{\rm i}2\Xi_{{\rm D}22}t}
\right)
+{\rm c.c.}
\right]\,,
\end{align}
where we have used that $\Pi_\chi^{\cal A}(k^0,\mathbf k)=-\Pi_\chi^{\cal A}(-k^0,\mathbf k)$, $\omega(\mathbf q)=\sqrt{\mathbf q^2+\bar M^2}$ and
we evaluate $\Pi^{\cal A}=\Pi^{\cal A}(\omega(\mathbf q),\mathbf q)$.
The parameters $\Delta$, $D$ and $\Xi$ are given by
Eqs.~(\ref{OmegaGamma:Scalar},\ref{diagonalisation:parameters})\,.

In order to assess the role of the oscillatory contributions within this
source term, we integrate
\begin{align}
\int\limits_0^\infty
dt\,
S_\varphi&=
-{\rm i}(g_1^2{g_2^*}^2-{g_1^*}^2g_2^2)
\int\frac{d^3 q}{(2\pi)^3 2\sqrt{{\mathbf q}^2+\bar M^2}}
\delta f_0(\mathbf q)
\frac{{\hat{\Pi}_\chi^{{\cal A}^2}}}{\left|(\Delta+D)^2-|\Pi^{\cal A}_{\chi_{12}}/\omega(\mathbf q)|^2\right|^2}
\\\notag
&\times
\Bigg[
|\Delta+D|^2(\Delta+D)
\left(
\frac1{\Gamma_{{\rm D}11}}
-\frac1{\frac{\rm i}2(\Omega_{{\rm D}11}-\Omega_{{\rm D}22})-\frac12(\Gamma_{{\rm D}11}-\Gamma_{{\rm D}22})}
\right)
\\\notag
&-
(\Delta^*+D^*)\left|\frac{\Pi^{\cal A}_{\chi_{12}}}{\omega(\mathbf q)}\right|^2
\left(
\frac1{\frac{\rm i}2(\Omega_{{\rm D}11}-\Omega_{{\rm D}22})-\frac12(\Gamma_{{\rm D}11}-\Gamma_{{\rm D}22})}
-\frac1{\Gamma_{{\rm D}22}}
\right)
+{\rm c.c.}
\Bigg]\,,
\end{align}
where we define $\Omega_{{\rm D}ii}={\rm Re}[\Xi_{{\rm D}{ii}}]$
and $\Gamma_{{\rm D}ii}=-{\rm Im}[\Xi_{{\rm D}{ii}}]$.
This result corresponds to the asymmetry from inverse
decays while $\chi_1$ approaches equilibrium, when
washout of the charge in $\varphi$ is ignored.
When $\Delta M\gg \Gamma_{\rm D}$, this becomes
\begin{align}
S_\varphi&=
-2{\rm i}(g_1^2{g_2^*}^2-{g_1^*}^2g_2^2)
\int\frac{d^3 q}{(2\pi)^3 2\sqrt{{\mathbf q}^2+\bar M^2}}
\delta f_0(\mathbf q)\frac{{\hat{\Pi}_\chi^{{\cal A}^2}}}{M^2_{\chi_{11}}-M^2_{\chi_{22}}}
\frac{\omega(\mathbf q)}{\Pi_{\chi_{11}}^{\cal A}}\,.
\end{align}
The same answer can be obtained from integrating the
source~(\ref{source:phi:hierarch}) for $\Gamma_{\rm D}\ll\Delta M$,
when substituting
$\delta f_{\chi_{ii}}=\delta f_0\exp(-(\Pi^{\cal A}_{\chi_{11}}(\omega(\mathbf q),\mathbf q)/\omega(\mathbf q))t)$.
We can therefore conclude that indeed, oscillatory contributions
to the $CP$-violating source approximately cancel in this regime.

Next, we compute the source~(\ref{source})
for the lepton asymmetry in a static background. For this purpose, we
need to calculate the relevant components of the singlet neutrino
propagator, $P_{\rm R}{\rm i}\delta S_{Nij}$. These can be expressed through the
distribution functions~(\ref{densities:analytical}), where the parameters
$\Delta$, $D$, $\Gamma$ follow from
Eqs.~(\ref{diagonalisation:parameters}) and~(\ref{Omega:Gamma:fermi}).
As we ignore the effect of washout, we also neglect the lepton and Higgs chemical
potentials when computing the singlet propagators and make use
of the approximation~(\ref{SigmaAN:simple}).
Notice that there are now two separate solutions for the helicity states
$h=\pm$, and we can therefore identify the solutions~(\ref{densities:analytical})
with $\delta f_{0h}$, {\it i.e.} $\delta f_{0hij}(k)=\delta f_{ij}(k)$. The
solutions thus following have the property
$\delta f_{0hij}(k^0,\mathbf k)=\delta f^*_{0hij}(-k^0,\mathbf k)=\delta f_{0hji}(-k^0,\mathbf k)$,
as a consequence of the Majorana condition~(\ref{Majorana:Propagator})
and the hermiticity~(\ref{hermiticity:fermi}) of
the distribution functions. Making use of this relation, we
can conveniently express the sum over positive and negative
frequencies that enters into the source for the asymmetry~(\ref{source})
as
\begin{align}
\label{source:decomposed}
{\cal S}(\mathbf k)
=
&
-Y_i^* Y_j
\int\frac{d k^0}{2\pi}
{\rm tr}
\left[
P_{\rm R}{\rm i}\delta S_{Nij}(k)2P_{\rm L}\hat{\slashed\Sigma}_N^{{\cal A}}(k)
\right]
\\\notag
=&Y_i^* Y_j\frac12
\sum\limits_{\underset{h=\pm}{k^0=\pm\omega(\mathbf k)}}
{\rm tr}\left[
P_{\rm R}(\gamma^0-h \hat k^i \gamma^i) \hat{\slashed\Sigma}^{\cal A}_N(k)
\right]
{\rm sign}(k^0)\delta f_{0hij}\left(1+\frac{h|\mathbf k|}{k^0}\right)
\\\notag
=&Y_i^* Y_j
\sum\limits_{h=\pm}
\Bigg\{
\hat \Sigma_N^{{\cal A}0}
\left[
\delta f_{0hij}-\delta f_{0hij}^*+\frac{h|\mathbf k|}{k^0}(\delta f_{0hij}+\delta f_{0hij}^*)
\right]
\\\nonumber
&\hskip0.96cm
-h\hat k^i\hat \Sigma_N^{{\cal A}i}
\left[
\delta f_{0hij}+\delta f_{0hij}^*+\frac{h|\mathbf k|}{k^0}(\delta f_{0hij}-\delta f_{0hij}^*)
\right]
\Bigg\}_{k^0=\omega(\mathbf k)}
\\\nonumber
=&Y_i^* Y_j \sum\limits_{h=\pm}
\Bigg\{ \frac{k\cdot \hat \Sigma_N^{{\cal A}}}{k^0}(\delta f_{0hij}-\delta f_{0hij}^*) + h \frac{\tilde k\cdot \hat \Sigma_N^{{\cal A}}}{k^0}(\delta f_{0hij}+\delta f_{0hij}^*)\Bigg\}_{k^0=\omega(\mathbf k)}
\,,
\end{align}
where we have used that
$\hat \Sigma_N^{{\cal A}0}(k^0,\mathbf k)=
\hat \Sigma_N^{{\cal A}0}(-k^0,\mathbf k)$ and
$\hat \Sigma_N^{{\cal A}i}(k^0,\mathbf k)=
-\hat \Sigma_N^{{\cal A}i}(-k^0,\mathbf k)$.
Above result for ${\cal S}(\mathbf k)$ should
be useful when $\delta f$ is calculated numerically,
{\it i.e.} for phenomenological studies of Leptogenesis in the Early
Universe, as we outline in Section~\ref{section:effective:theory}.

When we neglect the expansion, we can again use the analytic
solutions for damped flavour oscillations in
Eqs.~(\ref{densities:analytical}).
Using Eqs.~(\ref{Omega:Gamma:fermi})
and~(\ref{diagonalisation:parameters})
and defining
\begin{align}
\delta \hat f_{12}=\delta f_{12}/({\rm i}\Gamma_{12})\,,
\end{align}
we obtain the relations
\begin{subequations}
\label{factorisation:helicity}
\begin{align}
\delta f_{+0\,12}(\omega(\mathbf k),\mathbf k)
&=-\Xi_{12}(h=+,k^0=\omega(\mathbf k))\delta\hat f_{12}(\omega(\mathbf k),\mathbf k)\,,
\\
\delta f_{-0\,12}(\omega(\mathbf k),\mathbf k)
&=\Xi_{12}^*(h=+,k^0=\omega(\mathbf k))\delta\hat f_{12}(\omega(\mathbf k),\mathbf k)\,,
\\
\delta f_{+0\,12}(-\omega(\mathbf k),\mathbf k)
&=-\Xi_{12}^*(h=+,k^0=\omega(\mathbf k))\delta \hat f_{12}^*(\omega(\mathbf k),\mathbf k)\,,
\\
\delta f_{-0\,12}(-\omega(\mathbf k),\mathbf k)
&=\Xi_{12}(h=+,k^0=\omega(\mathbf k))\delta \hat f_{12}^*(\omega(\mathbf k),\mathbf k)\,.
\end{align}
\end{subequations}
Note that in the mass-diagonal basis, $\Xi_{ij}=-{\rm i}\Gamma_{ij}$, for $i\not=j$.
Above factorisation can easily be understood when noting that $|\Gamma_{12}|^2$,
as given by Eq.~(\ref{Omega:Gamma:fermi}),
does not depend on helicity. Moreover, from Eqs.~(\ref{diagonalisation:parameters})
and~(\ref{Omega:Gamma:fermi}), it follows that $\Xi_{\rm D}$ does not depend
on the helicity either. Therefore, the complete helicity dependence is
isolated within the explicit front
factor of $\Gamma_{12}$ in Eq.~(\ref{offdiag:analytical}).
Substituting Eqs.~(\ref{factorisation:helicity}) into
Eq.~(\ref{source:decomposed}), we obtain the
simple expression\footnote{Note that the summation over
$i,j=1,2$ that is understood in Eq.~(\ref{source:decomposed}) by the sum convention
is explicitly performed in the following equation.}
\begin{align}
\label{source:fermi:toy}
{\cal S}(\mathbf k)
={\rm i}({Y_1^*}^2Y_2^2-Y_1^2{Y_2^*}^2)
2g_w\frac{\bar M^2}{\mathbf k^2+\bar M^2}
\hat\Sigma^{\cal A}_{N\mu}\hat\Sigma_N^{{\cal A}\mu}
(\delta \hat f_{12}+ \delta \hat f_{12}^*)\,.
\end{align}
This is the source for the lepton asymmetry that results from
inverse decays of $N_1$ while it approaches the equilibrium
distribution in a background of constant temperature.
As a consistency check,
the known result for $\Delta M\ll\bar M$ but
$\Delta M\gg\Gamma_{\rm D}$ is recovered for
$\delta \hat f_{12}=\delta f_0[\omega(\mathbf k)/(M_{11}^2-M_{22}^2)]\exp(-\Gamma_{11}t)$,
which results from
neglecting the oscillatory
contributions in Eq.~(\ref{offdiag:analytical}).

\begin{figure}[t!]
\begin{center}
\epsfig{file=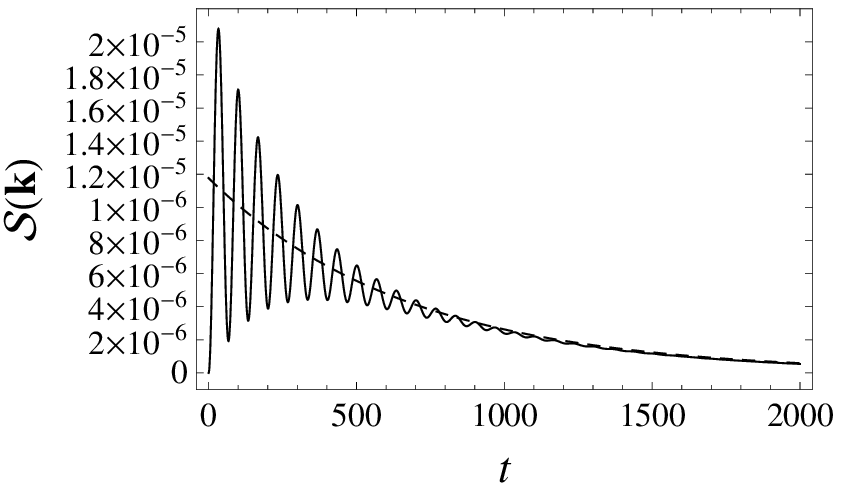,width=7.5cm}
\epsfig{file=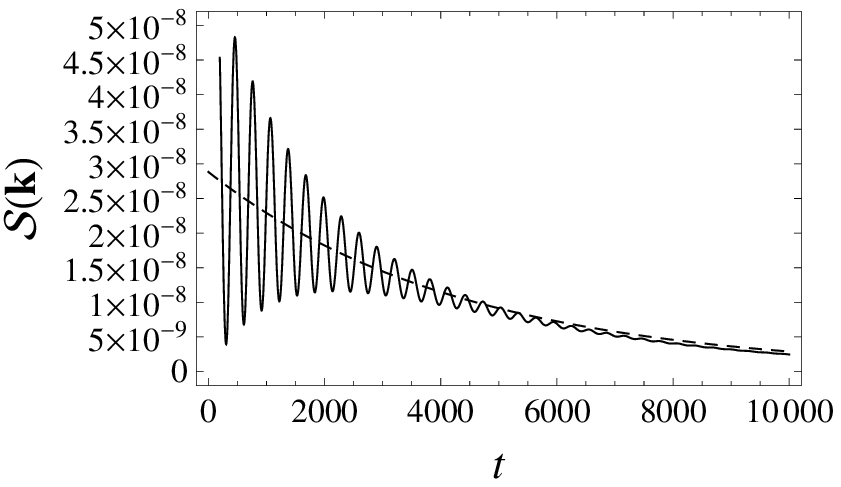,width=7.5cm}
\end{center}
\caption{
\label{fig:nonres}
The source ${\cal S}(\mathbf k)$ for the lepton asymmetry
over time $t$ for
$M_1=1$, $M_2=1.1$, $T=1$, $Y_1=0.1$, $Y_2=0.2+0.1{\rm i}$,
{\it i.e.} $\Delta M\gg \Gamma_{\rm D}$,
and initially a vanishing distribution for $N_1$ and
an equilibrium distribution for $N_2$. The solid line is
the result~(\ref{source:decomposed}), the dashed line
the standard perturbative limit~(\ref{diff:source:pert}).
In the left panel, we take $|\mathbf k|=0.5$, and consider the asymmetry
resulting from a non-relativistic singlet neutrino. In the
right panel, the singlet neutrino is relativistic, $|\mathbf k|=5$.
}
\end{figure}

\begin{figure}[t!]
\begin{center}
\epsfig{file=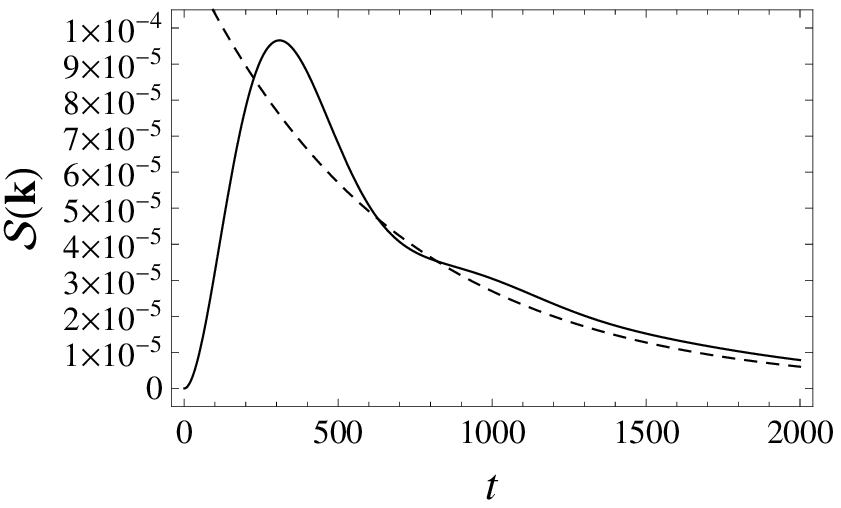,width=7.5cm}
\epsfig{file=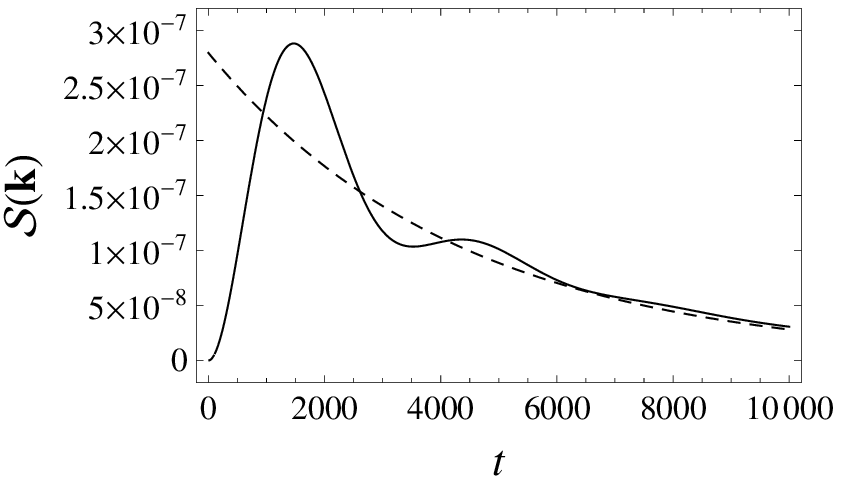,width=7.5cm}
\end{center}
\caption{
\label{fig:res}
The source ${\cal S}(\mathbf k)$ over time $t$
for the lepton asymmetry for
$M_1=1$, $M_2=1.01$, $T=1$, $Y_1=0.1$, $Y_2=0.2+0.1{\rm i}$,
{\it i.e.} $\Delta M\sim \Gamma_{\rm D}$,
and initially a vanishing distribution for $N_1$ and
an equilibrium distribution for $N_2$. The solid line is
the result~(\ref{source:decomposed}), the dashed line
the standard perturbative limit~(\ref{diff:source:pert}).
In the left panel, we take $|\mathbf k|=0.5$, and consider the asymmetry
resulting from a non-relativistic singlet neutrino. In the
right panel, the singlet neutrino is relativistic, $|\mathbf k|=5$.
}
\end{figure}

We now compare the result from Eqs.~(\ref{source:decomposed})
or~(\ref{source:fermi:toy}) to the standard
perturbative approximation
\begin{align}
\label{diff:source:pert}
{\cal S}(\mathbf k)
=\frac{8{\rm i}[{Y_1^*}^2Y_2^2-Y_1^2{Y_2^*}^2]}{2\sqrt{\mathbf k^2+\bar M^2}}
\frac{M_{11}M_{22}}{M_{11}^2-M_{22}^2}g_w\hat\Sigma^{\cal A}_{N\mu}\hat\Sigma_N^{{\cal A}\mu}
{\rm e}^{-2g_w|Y_1|^2(k^\mu\hat\Sigma^{\cal A}_{N\mu}/k^0)t}\delta f_0
\,.
\end{align}
For this purpose,
we use the solutions~(\ref{densities:analytical}) with
$\delta f_0(\mathbf k)=-f^{\rm eq}(\mathbf k)$,
which corresponds to a vanishing number of $N_1$ at $t=0$
in the
particular mode $\mathbf k$, while $N_2$ is in equilibrium.
The other parameters are given in the Figure captions. Their choice
does not correspond to a systematic study of parameter space,
but the examples should be illustrative and representative.
In Figure~\ref{fig:nonres}, we illustrate the situation when
$\Delta M\gg\Gamma_{\rm D}$. The oscillation frequency is much
larger than the damping rate, such that we may expect that
these oscillations average out in the final result for the asymmetry,
see the discussion above. We choose examples of non-relativistic and
relativistic singlet neutrinos, because in the relativistic regime,
the result turns out to be the remainder of an incomplete
cancellation of larger terms involving
$\hat{\Sigma}^{{\cal A}0}_N$ and
$\hat{\Sigma}^{{\cal A}i}_N$, which is therefore explicitly verified.
The results in
Figure~\ref{fig:res} correspond to a point in parameter space where
$\Delta M\sim\Gamma_{\rm D}$. Since the damping rate is now of the
same order as the frequency of flavour oscillations between $N_1$
and $N_2$, these should not be neglected when calculating the
asymmetry.

This brings us to comment on earlier expressions on the behaviour
of the asymmetry when approaching the resonant regime. This matter
is discussed in Refs.~\cite{Buchmuller:1997yu,Pilaftsis:1997jf,Pilaftsis:2003gt,Anisimov:2005hr}, where expressions
for the asymmetries
are given. Oscillations are neglected there, an approximation that
is not always suitable, {\it cf.} Figure~\ref{fig:res}.
It is however emphasised in Refs.~\cite{Buchmuller:1997yu,Pilaftsis:1997jf,Pilaftsis:2003gt,Anisimov:2005hr}, that in the regimes
$\Delta M\sim\Gamma_{\rm D}$ and $\Delta M\ll\Gamma_{\rm D}$, the
standard perturbative result breaks down and a suitable resummation
technique for the singlet neutrino propagator should be employed,
which is essentially what is performed within the present work.
It has already been mentioned in Section~\ref{section:constant:background}, that when choosing
$U\delta f(0)U^\dagger$ as a diagonal matrix, there
are no flavour oscillations. It should however be emphasised
that this corresponds to a peculiar initial state that
already bears a $CP$-asymmetry.
From Eqs.~(\ref{diagonalisation:parameters}) and~(\ref{source:fermi:toy})
it then follows that
\begin{align}
{\cal S}(\mathbf k)
={\rm i}({Y_1^*}^2Y_2^2-Y_1^2{Y_2^*}^2)
2g_w\frac{\bar M}{\mathbf k^2+\bar M^2}
\hat\Sigma^{\cal A}_{N\mu}\hat\Sigma_N^{{\cal A}\mu}
\left(
\frac{1}{\Delta +D}+\frac{1}{\Delta^*+D^*}
\right)\delta f_{0hii}
\,.
\end{align}
Notice that $\Delta+D\not=0$, as long as $Y_i\not=0$.
This result is in agreement with Refs.~\cite{Liu:1993ds,Covi:1996fm},
where the time evolution of mixing scalars in the vacuum is considered.
It generalises the findings of Refs.~\cite{Liu:1993ds,Covi:1996fm}
to mixing Majorana fermions, that decay in a finite-temperature
background.
In the limit $\Delta M\gg\Gamma_{\rm D}$, the term
$\sim\Xi_{12}\Xi_{21}$ in the discriminant of $D$ can be neglected,
and the results of Refs.~\cite{Buchmuller:1997yu,Anisimov:2005hr} 
for Majorana neutrino decay in a zero temperature
background is recovered. We emphasise however, that when
the condition $\Delta M\gg\Gamma_{\rm D}$ does not apply, the
full time-evolution of the off-diagonal components of the singlet
neutrino propagator has to be calculated,
because due to oscillation effects, these are not simply proportional to
the diagonal components in general.

Finally, let us briefly discuss the limit $|\Gamma_{12}| \ll |\Gamma_{11} - \Gamma_{22}|$ and $|\Gamma_{12}| \lesssim \Delta M$ in order to facilitate the comparison with the results in Ref.~\cite{Garny:2011}. In this limit the term
$\sim\Xi_{12}\Xi_{21}$ in the discriminant of $D$ can again be neglected, and Eq.~(\ref{source:fermi:toy}) for the $CP$-violating source reduces to
\begin{align}
{\cal S}(\mathbf k)
=&\frac{8{\rm i}[{Y_1^*}^2Y_2^2-Y_1^2{Y_2^*}^2]}{2\sqrt{\mathbf k^2+\bar M^2}}
\frac{M_{11}M_{22}(M_{11}^2-M_{22}^2)}{(M_{11}^2-M_{22}^2)^2 + \omega^2(\Gamma_{11} - \Gamma_{22})^2}g_w\hat\Sigma^{\cal A}_{N\mu}\hat\Sigma_N^{{\cal A}\mu}
\notag\\
&\times\bigg[{\rm e}^{-\Gamma_{11}t} - \Big( \cos(\Delta\omega t) + \frac{\Gamma_{11}-\Gamma_{22}}{2\Delta\omega}\sin(\Delta\omega t)\Big){\rm e}^{-(\Gamma_{11}+\Gamma_{22})t/2}\bigg]\delta f_0
\,.
\end{align}
where $\Delta\omega(\mathbf k) \equiv |M_{11}^2 - M_{22}^2|/(2\omega(\mathbf k))$.
This result is found to be in agreement with the analytical results presented in Ref.~\cite{Garny:2011}.\footnote{Note that in Ref.~\cite{Garny:2011} the initial state corresponds to vanishing number densities for {\em both} neutrino flavours $N_1$ and $N_2$. Furthermore, the analytical results in \cite{Garny:2011} assume the non-relativistic limit $M_i \gg T$, such that $\hat\Sigma^{\cal A}_{N\mu}(k) \approx k^\mu/(32\pi)$.} Furthermore, the structure of the denominator $(M_{11}^2-M_{22}^2)^2 + \omega^2(\Gamma_{11} - \Gamma_{22})^2$ agrees with the results in ref.~\cite{Anisimov:2005hr}. However, we want to emphasize that the parametric regime with $|\Gamma_{12}| \ll |\Gamma_{11} - \Gamma_{22}|$ is by no means generic in resonant leptogenesis.

\section{Effective Theory in the Expanding Universe}
\label{section:effective:theory}

In this Section, we present a network of equations that describes the 
simplest version of resonant Leptogenesis in the expanding Universe.
Equations that are valid sufficiently far
away from the resonant limit, {\it i.e.} for $\Delta M\gg \Gamma_{\rm D}$,
are presented in Ref.~\cite{Beneke:2010wd}.
In the present context, we need to generalise Eq.~(\ref{flavour:dyn:fermi})
in such a way, that it encompasses the effect of the expanding background,
in particular how this induces a deviation of the distribution functions
of the singlet neutrinos from equilibrium.

First, we transform the time coordinate $t$ to the conformal time $\eta$
through the relation $dt=a(\eta)d\eta$, where $a(\eta)$ denotes the scale
factor. In the radiation dominated Universe, $a(\eta)=a_{\rm R}\eta$, where
$a_{\rm R}$ is a constant. The scale factor is related to the
physical temperature by
\begin{align}
T=\frac{1}{a_{\rm R}\eta}\sqrt{\frac{a_{\rm R}m_{\rm Pl}}{2}}
\left(\frac{45}{\pi^3 g_*}\right)^{1/4}\,,
\end{align}
where $g_*$ denotes the number of relativistic degrees of freedom.
Note that the comoving temperature $T_{\rm com}=a(\eta)T$ is constant.
There
remains a freedom of parametrisation for $a_{\rm R}$. For kinetic equations
in the early Universe, $z=\bar M\eta$ is a convenient choice, where
\begin{align}
z=\frac{\bar M}{T}\,,
\end{align}
and therefore
\begin{align}
\label{aR:z:eta}
a_{\rm R}=\frac{m_{\rm Pl}}{2}\sqrt{\frac{45}{\pi^3g_*}}\,.
\end{align}
The choice of the conformal time $\eta$ instead of the comoving proper time
$t$ has the advantage that the Kadanoff Baym equations are
of a form that is very similar to the one they take in Minkowski background.
The only modification is that the mass terms need to be rescaled as
$M\to a(\eta)  M$ and the momenta are to be understood as comoving
momenta. The derivative with respect to $\eta$ [or, equivalently, $z/\bar M$, when
using Eq.~(\ref{aR:z:eta})] is denoted by a prime.

Using these parametrisations, one can obtain a term that represents
the deviation of the singlet neutrino from equilibrium, that is induced
by the expansion of the Universe. When $\Delta M\ll\bar M$, the
equilibrium distribution for the singlet neutrinos is
\begin{align}
f^{\rm eq}_{Nii}(\eta)
=\frac{1}{{\rm e}^{\sqrt{\mathbf k^2+a^2(\eta) \bar M^2}/T_{\rm com}}+1}
\end{align}
and $f^{\rm eq}_{Nij}(z)=0$ for $i\not=j$. When we are not working
in the mass eigenbasis, this relation is only approximately correct and 
there will be off-diagonal elements that are suppressed by a factor
of $\Delta M/M$ when compared to the diagonal ones. While this is
a tolerable error, it could easily be corrected for by diagonalising
$M$ through an orthogonal transformation.
The expanding background induces a decrease of the temperature,
such that consequently,
\begin{align}
\frac{d}{d\eta}f^{\rm eq}_{Nii}(\eta)
=f^{{\rm eq}\prime}_{Nii}(\eta)
=-
\frac{{\rm e}^{\sqrt{\mathbf k^2+a^2(\eta) \bar M^2}/T_{\rm com}}}
{\left({\rm e}^{\sqrt{\mathbf k^2+a^2(\eta) \bar M^2}/T_{\rm com}}+1\right)^2}
\frac{a_{\rm R}^2\bar M^2\eta}
{T_{\rm com}\sqrt{\mathbf k^2+a^2(\eta) \bar M^2}}
\,,
\end{align}
and $f^{{\rm eq}\prime}_{Nij}(z)=0$ for $i\not=j$.

The generalisation of Eq.~(\ref{flavour:dyn:fermi}), valid in Minkowski
background, to an expanding, radiation dominated Universe is now given
by
\begin{align}
\label{evol:neutrino}
\delta f_{0h}^\prime
+a^2(\eta)\frac{1}{2k^0}
{\rm i}[M^2,\delta f_{0h}]
+f^{{\rm eq}\prime}
=&-
\left(\Sigma_{N{\rm L}}^{{\cal A}0}
+h\hat k^i\Sigma_{N{\rm L}}^{{\cal A}i}\right)
\left(
1-\frac{h|\mathbf k|}{k^0}
\right)
\left\{\mathbf Y^*\mathbf Y^t,
\delta f_{0h}
\right\}
\\\notag
&-
\left(
\Sigma_{N{\rm R}}^{{\cal A}0}
-h\hat k^i\Sigma_{N{\rm R}}^{{\cal A}i}
\right)
\left(
1+\frac{h|\mathbf k|}{k^0}
\right)
\left\{
\mathbf Y\mathbf Y^\dagger,
\delta f_{0h}
\right\}\,,
\\\notag
\textnormal{where}\quad
k^0=&\pm\sqrt{\mathbf k^2+a(\eta)^2\bar M^2}\,.
\end{align}
Note that this equation makes no assumption about the form
of $M$, except that it is a real symmetric matrix, which allows
to account for temperature-dependent effective mass-corrections that
may be incurred through $\slashed\Sigma_N^H$.
Eq.~(\ref{evol:neutrino}) is therefore flavour-covariant with respect to
orthogonal transformations of $N_i$. Allowing for a complex symmetric
$M$ would lead to a more complicated form of Eq.~(\ref{evol:neutrino}),
but it would not capture addtional physical parameter space,
as the complex phases in $M$ can always be absorbed in $Y$,
{\it cf.} the discussion in Section~\ref{section:CTPLepto}. Therefore,
Eq.~(\ref{evol:neutrino}) can be used also in situations where the mass eigenbasis is
time-dependent due to finite-temperature effects.
This should be of importance close to the
resonant limit, since the thermal corrections to $M$ that arise from
$\slashed\Sigma_N^H$ are given by (for momenta $|\mathbf k|\gg|M_{ij}^{\rm Th}|$,
which covers the relevant contributions of phase-space)
\begin{align}
\label{eff:mass:square}
{M_{ij}^{{\rm Th}}}^2
=\frac14\left(Y_i^*Y_j+Y_iY_j^*\right)T^2
\,,
\end{align}
which is of the same order as $k^0\Gamma_{\rm D}$.

Assuming that the lepton charge density is small compared to
the number density (or, equivalently, the lepton chemical potential
$\mu_\ell$ is small compared to the temperature $T$), we extract
from Eq.~(\ref{source:washout}) the washout term as
\begin{align}
\label{washout}
W=&Y_i^* Y_j \int\frac{d^4k}{(2\pi)^4}
\int\frac{d^4k^\prime}{(2\pi)^4}
\int\frac{d^4k^{\prime\prime}}{(2\pi)^4}
(2\pi)^4\delta^4(k-k^\prime-k^{\prime\prime})
\\\notag
\times&
{\rm tr}
\left[
\left(
P_{\rm R}
{\rm i}S_{Nij}^>(k^\prime)
P_{\rm L}
{\rm i}\Delta_\phi^<(-k^{\prime\prime})
-
P_{\rm R}
{\rm i}S_{Nij}^<(k^\prime)
P_{\rm L}
{\rm i}\Delta_\phi^>(-k^{\prime\prime})
\right)
{\rm i}\delta S_{\ell}(k)
\right]
\,.
\end{align}
The deviation of the lepton propagator 
${\rm i}\delta S_{\ell}(k)$ from the form with vanishing
chemical potential can be expressed through
\begin{align}
\label{lepto:chempot}
\delta f_\ell(\mathbf k)=-\delta \bar f_\ell(\mathbf k)
=(n_\ell-\bar n_\ell)\times
\frac{6 {\rm e}^{|\mathbf k|/T}}{T^3\left({\rm e}^{|\mathbf k|/T}+1\right)^2}
=\mu_\ell\frac{{\rm e}^{|\mathbf k|/T}}{T\left({\rm e}^{|\mathbf k|/T}+1\right)^2}
\,,
\end{align}
where
\begin{align}
{\rm i}\delta S_\ell(k)
=-P_{\rm L}\slashed k
2\pi\delta(k^2)
\delta f_\ell(\mathbf k){\rm sign}(k^0)
\,.
\end{align}
In contrast to the washout term quoted in Ref.~\cite{Beneke:2010wd},
we include here the off-diagonal elements of the singlet neutrino
propagator. When $\Delta M \gg \Gamma_{\rm D}$, these are
suppressed compared to the out-of-equilibrium
components by a factor of $\Gamma_{\rm D}/\Delta M$, {\it i.e.}
we can safely neglect these in the non-resonant regime. When
$\Delta M$ is of order $\Gamma_{\rm D}$ or smaller, neglecting
the off-diagonal correlations is not necessarily a good approximation.
Especially in the weak washout regime $\delta f_{0h}$ may
be of similar size as $f^{\rm eq}$, and these corrections
may be of importance. In the strong washout regime however, the
deviation of the singlet neutrino density from equilibrium is small,
such that we can substitute the equilibrium propagator for
${\rm i}S_N$ in Eq.~(\ref{washout}), which is diagonal
in the mass eigenbasis, when neglecting finite-width contributions.

The collision term on the right-hand-side of Eq.~(\ref{evol:neutrino}) is valid when the chemical potentials
of the Higgs and lepton fields can be neglected. Once this
is not the case, it should be replaced with
\begin{align}
&-\frac{\rm i}2
\left(\Sigma_{N{\rm L}}^{>0}
+h\hat k^i\Sigma_{N{\rm L}}^{>i}\right)
\left(
1-\frac{h|\mathbf k|}{k^0}
\right)
\left\{\mathbf Y^*\mathbf Y^t,
f_{0h}
\right\}
\\\notag
&-\frac{\rm i}2
\left(
\Sigma_{N{\rm R}}^{>0}
-h\hat k^i\Sigma_{N{\rm R}}^{>i}
\right)
\left(
1+\frac{h|\mathbf k|}{k^0}
\right)
\left\{
\mathbf Y\mathbf Y^\dagger,
f_{0h}
\right\}
\\\notag
&+\frac{\rm i}2
\left(\Sigma_{N{\rm L}}^{<0}
+h\hat k^i\Sigma_{N{\rm L}}^{<i}\right)
\left(
1-\frac{h|\mathbf k|}{k^0}
\right)
\left\{\mathbf Y^*\mathbf Y^t,
1+ f_{0h}
\right\}
\\\notag
&+\frac{\rm i}2
\left(
\Sigma_{N{\rm R}}^{<0}
-h\hat k^i\Sigma_{N{\rm R}}^{<i}
\right)
\left(
1+\frac{h|\mathbf k|}{k^0}
\right)
\left\{
\mathbf Y\mathbf Y^\dagger,
1+f_{0h}
\right\}\,.
\end{align}
Here, $f_{0h}=f^{\rm eq}_{0h}+\delta f_{0h}$ is the
full distribution matrix of $N$,
that also includes the equilibrium distribution
$f^{\rm eq}_{0h}(\mathbf k)=
{\rm diag}(1/(\exp(\sqrt{\mathbf k^2+M_{11}^2}/T)+1),1/(\exp(\sqrt{\mathbf k^2+M_{22}^2}/T)+1))$.
The $\Sigma_N^{<,>}$ can be obtained from
the expressions for $\Sigma_N^{\cal A}$
in Appendix~\ref{appendix:selfergs} when
using the Kubo-Martin-Schwinger (KMS) relations
\begin{align}
\Sigma_{N{\rm L,R}}^{>}(k)=
-{\rm e}^{(k^0\mp\mu_\ell\mp\mu_\phi)/T}\Sigma_{N{\rm L,R}}^{>}(k)\,.
\end{align}
It should be useful to expand this term in the small ratios
$\mu_{\ell,\phi}/T$, in order to separate it into a
contribution of the form as in Eq.~(\ref{evol:neutrino}) and an
additional one, that only depends on the chemical potentials.
A detailed study of this matter is subject of ongoing work.

Of course, Eq.~(\ref{source:washout}) must be transformed from
comoving to conformal time as well, with the result
\begin{align}
\label{evol:lepto}
\frac{d}{d\eta}(n_\ell-\bar n_\ell)=W+S\,.
\end{align}
For $\Delta M\gg \Gamma_{\rm D}$,
the procedure of solving for the lepton asymmetry is described
in detail in  Ref.~\cite{Beneke:2010wd}.
We first determine $\delta f_{0hij}(k)$ from
Eq.~(\ref{evol:neutrino}) and the approximation~(\ref{SigmaAN:simple})
of vanishing chemical potentials for Higgs and leptons.
The result is then to be substituted
into the source term~(\ref{source}) and the washout
term~(\ref{washout}).
When $\Delta M \lesssim \Gamma_{\rm D}$, we should not expect
the approximation~(\ref{SigmaAN:simple}) substituted
into the kinetic equations for the oscillating singlet
neutrinos~(\ref{evol:neutrino})
to be sufficient,
as it may fail to accurately describe the washout of the lepton
asymmetry. In that situation, the evolution of the
lepton asymmetry~(\ref{evol:lepto}) and of the correlation
of the singlet neutrinos~(\ref{evol:neutrino}) should be solved
as a coupled set of equations. In principle, the lepton and
the Higgs chemical potentials are further processed by spectator effects.
A good approximation for first phenomenological studies may
be to consider an unflavoured regime, where the charge
in $\ell$ is conserved, up to the interactions with $N_i$, that
are accounted for already
by our equations. The hypercharge asymmetry within $\phi$
should get distributed over a large number of remaining Standard Model
degrees of freedom, such that we may set $\mu_\phi\approx 0$ within a
first approximation.

Note that for the source term,
the form given by Eq.~(\ref{source:decomposed}) may
be particularly useful,
because
we only have to calculate the positive frequency solutions for $\delta f_{0hij}(k)$, while the negative frequency solution follows as a 
consequence of the Majorana condition.
Eventually, Eqs.~(\ref{evol:neutrino}) and~(\ref{evol:lepto}) can be
integrated in order to obtain the final lepton asymmetry.
A numerical study of the evolution of the lepton asymmetry,
using the methods that are outlined in this Section, should
be performed in future work.
In particular, when $\Delta M \gg \Gamma_{\rm D}$,
it appears plausible that for an equilibrium deviation induced by
the expanding background, the oscillating
contributions visible in Figure~\ref{fig:nonres} vanish
altogether, because the deviation of the singlet neutrinos
from equilibrium is generated continuously throughout Leptogenesis.
Deviations that appear at different times oscillate and have different
phases, such that the oscillations may effectively vanish due to
decoherence. However, when $\Delta M\approx\Gamma_{\rm D}$ or
$\Delta M\ll\Gamma_{\rm D}$, the methods outlined in this
Section may be suitable to make predictions for resonant Leptogenesis
that are not accurately covered by the perturbative
result~(\ref{source:well-known}) for the source term.

\section{Summary and Conclusions}
\label{section:conclusions}

In this paper, we have developed the theory of Leptogenesis
from mixing Majorana fermions and from mixing scalars in the
CTP formalism. Our calculations rely on the expansion
parameters
$\Gamma_{\rm D}\ll \Delta M$ or $\Delta M\ll \bar M$. Due to
the smallness of the couplings, $\Gamma_{\rm D}\ll\bar M$,
such that the parametric regimes where the two approximations apply
overlap. When $\Gamma_{\rm D}\ll \Delta M$, the usual
result for the $CP$-asymmetry from the wave-function correction
applies~\cite{Flanz:1996fb,Covi:1996wh,Pilaftsis:1997dr,Pilaftsis:1997jf,Pilaftsis:2003gt,Pilaftsis:2005rv,Beneke:2010wd}.
Otherwise, oscillations are relevant for the $CP$ asymmetry, and it
is necessary to solve the kinetic equations for
the singlet neutrino propagator.
In Section~\ref{section:effective:theory}, we explain how such a solution
may be obtained when taking the expansion of the Universe into account
and how this solution can be used to calculate the final value
of the lepton asymmetry of the Universe.

Within our results, we recover many elements
that have previously been derived using Boltzmann equations
and $S$-matrix elements or the
Hamiltonian
approach. However, there are a few features that differ from
the earlier approaches:
\begin{itemize}
\item
For Leptogenesis from the decays of Majorana fermions,
we show that the kinetic evolution
equations~(\ref{flavour:dynamic:general}) can accurately
reproduce the classic result~\cite{Flanz:1996fb,Covi:1996wh,Pilaftsis:1997dr,Pilaftsis:1997jf,Pilaftsis:2003gt,Pilaftsis:2005rv,Beneke:2010wd} for the asymmetry in Resonant
Leptogenesis~(\ref{source:well-known}). This requires a detailed
analytical computation of the evolution of the two helicity
states of the singlet neutrinos, and has not been provided
in earlier work based on the Hamiltonian
approach~\cite{Akhmedov:1998qx,Asaka:2005pn,Gagnon:2010kt}.
\item
Our results consistently
include the quantum statistical (Fermi-Dirac and Bose-Einstein)
corrections that appear for all on-shell particles in the
finite-density background. Note that these also apply to
the on-shell particles within the loops of the diagrams that
describe $CP$-violation.
\item
When $\Delta M \sim \Gamma_{\rm D}$ or smaller, it is important
to take account of the lepton and Higgs chemical potentials, because
in this situation, the lepton number violation by the singlet neutrinos
proceeds so slowly, that we have to take account of its time-dependence.
This may be crucial in order to treat the washout of the asymmetry
in a quantitatively correct manner.
\item
Our effective theory is formulated within one single framework, the CTP 
formalism. In contrast, earlier approaches typically combine elements
from $S$-matrices, classical Boltzmann equations or Hamiltonian
transition amplitudes. In particular, in combination with
Ref.~\cite{Beneke:2010dz}, the methods presented here cover both, direct
$CP$-violation 
and $CP$-violation from mixing, even when oscillations are important.
Working within a single formalism may have
the advantage, that it is easier to implement controlled approximations.
The CTP approach combines the advantage of the $S$-matrix approach, that
perturbative expansions can be efficiently formulated in terms of Feynman 
diagrams, with the advantage of the Hamiltonian real-time evolution,
that avoids unitarity violation from the overcounting of real intermediate
states.
\item
Since the CTP approach is formulated in terms of Green functions
and Feynman diagrams,
this method for calculating the lepton asymmetry
may be particularly suitable to be combined with
corrections due to scattering processes in the thermal
bath~\cite{Asaka:2006rw,Kiessig:2010pr,Anisimov:2010gy,Salvio:2011sf,Kiessig:2011fw,Kiessig:2011ga,Laine:2011pq}.
\end{itemize}

The CTP approach appears as a well-suited method for a consistent
description of macroscopic statistical systems that consist of
High Energy Particle Physics degrees of freedom. In particular, it
may provide an intuitive and accurate framework in order
to describe the emergence of the baryon asymmetry of the Universe
from the quantum effect of $CP$-violation.
Further significant progress on the theory of Leptogenesis appears to
be a realistic prospect. For Resonant Leptogenesis,
the present work is intended to pave the
way to improve the quantitative predictions and the qualitative
understanding.

\subsubsection*{Additional Notes}
Another successful calculation of the asymmetry for Resonant Leptogenesis in
the CTP approach, using partly different calculational strategies than those
in the present work, was reported by Garny et al.~\cite{Garny:2011}.
During the final stage of the preparation of this manuscript,
Ref.~\cite{Asaka:2011wq} appeared. Based on the Hamiltonian approach, it is
also pointed out there, that the lepton chemical potential is relevant for
computing the correlations of the singlet neutrino flavours.

\subsubsection*{Acknowledgements}

\noindent
This work is supported by the Gottfried Wilhelm Leibniz programme
of the Deutsche Forschungsgemeinschaft and by the
Alexander von Humboldt Foundation.

\begin{appendix}

\section{Two-Point Functions on the CTP}

\label{appendix:CTP:Wigner}

Two-point functions on the CTP are endowed with two
path indices $+/-$. We often make use of the definitions
\begin{subequations}
\begin{align}
G^{++}(x,y)&=G^T(x,y)\,,\\
G^{+-}(x,y)&=G^<(x,y)\,,\\
G^{-+}(x,y)&=G^>(x,y)\,,\\
G^{--}(x,y)&=G^{\bar T}(x,y)\,,
\end{align}
\end{subequations}
where $G$ stands for any two-point function, {\it i.e.} fermionic
or scalar, or Green function or propagator.

The propagators of a scalar field $\chi$ are
\begin{subequations}
\begin{align}
  {\rm i}\Delta^T(x,y) 
   &= \langle
        T [\chi(x) \chi^\dagger(y) ]
     \rangle\,, 
\\
  {\rm i}\Delta^<(u,v)
   &= \langle
        \chi^\dagger(y) \chi(x) 
     \rangle\,,   
\\
  {\rm i}\Delta^>(x,y)
   &= \langle
        \chi(x) \chi^\dagger(y) 
     \rangle\,, 
\\
  {\rm i}\Delta^{\bar T}(x,y)
   &= \langle
        \overline T [ \chi(x) \chi^\dagger(y) ] 
     \rangle   
\,.
\end{align}
\end{subequations}
In order to describe mixing of flavours, $\chi$
ought to be considered as a column vector in flavour space.
$T$ ($\bar T$) stands for (anti-)time ordering and the superscript
for the corresponding Green functions. We sometimes refer to the Green
functions with the superscripts $<,>$ as Wightman functions.
When $\chi$ is real, the neutrality constraint
$\chi^c=\chi^\dagger=\chi$ must be observed.
It then follows that
\begin{align}
\label{neutrality:scalar}
\Delta(x,y)=\Delta^t(y,x)\,,
\end{align}
where the superscript $t$ denotes the transposition
which acts here on both, CTP ($\pm$) and flavour indices.
For both, real and complex fields the Wightman functions
have the hermiticity property
\begin{align}
\label{hermiticity:scalar}
{\rm i}\Delta^{<,>}(x,y)=\left({\rm i}\Delta^{<,>}(y,x)\right)^\dagger.
\end{align}

Similarly, for a fermion $\psi$, the Green functions are
\begin{subequations}
\begin{align}
  {\rm i}S^T(x,y)
   &=   \langle
                T [ \psi(x) \bar{\psi}(y)]
             \rangle\,,   
\\
  {\rm i}S^<(x,y)
   &= -       \langle
                \bar{\psi}(y) \psi(x)
             \rangle\,,   
\\
  {\rm i}S^>(x,y)
   &=  \langle
              \psi(x) \bar{\psi}(y) 
             \rangle\,,   
\\
  {\rm i}S^{\bar{T}}(x,y)
   &=  \langle
                \overline{T} [\psi(x) \bar{\psi}(y)]
              \rangle
\,.
\end{align}
\end{subequations}
Notice that these define matrices in Spinor space.
A Majorana spinor mus observe the condition
\begin{align}
\label{Majorana:condition}
\psi^c=C \bar \psi^t=\psi\,,
\end{align}
where $C$ is the charge conjugation matrix.
As a consequence, for Majorana two-point functions,
\begin{align}
\label{Majorana:Propagator}
S(x,y)=C S^t(y,x) C^\dagger\,,
\end{align}
where the transposition acts on all, CTP ($\pm$), spinor and flavour indices. The hermiticity property
of the Wightman function reads now
\begin{align}
\label{hermiticity:fermi}
{\rm i}\gamma^0 S^{<,>}(x,y)=\left({\rm i}\gamma^0 S^{<,>}(y,x)\right)^\dagger.
\end{align}
which holds for both, charged and Majorana spinors.

Additional useful two-point functions follow when taking the
combinations
\begin{subequations}
\label{CTP:combinations}
\begin{align}
\label{CTP:advanced}
G^A=G^T-G^>=G^<-G^{\bar T}\quad&\textnormal{(advanced)}\,,
\\
\label{CTP:retarded}
G^R=G^T-G^<=G^>-G^{\bar T}\quad&\textnormal{(retarded)}\,,
\\
\label{CTP:hermitian}
G^H=\frac12(G^R+G^A)\quad&\textnormal{(Hermitian)}\,,
\\
\label{CTP:spectral}
G^{\cal A}=\frac1{2\rm i}(G^A-G^R)=\frac{\rm i}2(G^>-G^<)\quad&\textnormal{(anti-Hermitian, spectral)}\,.
\end{align}
\end{subequations}

The Wigner transform of a two-point function is defined as
\begin{align}
\label{Wigner:transform}
G(k,x)=\int d^4 r {\rm e}^{{\rm i}kr} G(x+r/2,x-r/2)\,,
\end{align}
where we refer to $r$ as the relative and to $x$ as the average coordinate.
In the situation of spatial homogeneity, there is no dependence on $\mathbf x$
(spatial translation invariance), such that $G(k,x)\equiv G(k,t)$.

The transformation of convolution integrals in coordinate space
into Wigner space can be expressed with the help of
the diamond operator, which is defined as
\begin{align}
\label{diamond:operator}
\diamond\{A\}\{B\}=
\frac12\left(\partial_x A\right)\left(\partial_k B\right)
-\frac12\left(\partial_k A\right)\left(\partial_x B\right)\,,
\end{align}
where $A\equiv A(k,x)$ and $B\equiv B(k,x)$.

\section{Equilibrium Self-Energies}
\label{appendix:selfergs}

The spectral self-energy, that is central in calculations of
the resonantly generated lepton asymmetry is
computed in Ref.~\cite{Beneke:2010wd}.
Here, we generalise that expression in order to account for
the lepton and Higgs chemical potentials. This may be of importance
when $\Delta M \gg \Gamma_{\rm D}$ is not valid, such that lepton
number violation is modulated by the slow oscillations of the
right-handed neutrinos.
Neglecting tree-level and
thermal masses for the lepton and the Higgs boson, the
normalised spectral self-energy can be expressed as
\begin{subequations}
\label{Sigma:analytic}
\begin{align}
\hat\Sigma^{{\cal A}0}_{N{\rm L,R}}(k) &= \frac{T^2}{16 \pi |{\mathbf k}|}
\left[
\vartheta(k^0)
I_1\left(\frac{|k^0|}{T},\frac{|\mathbf k|}{T},\mp\frac{\mu_\ell}{T},\mp\frac{\mu_\phi}{T}\right)
+\vartheta(-k^0)
I_1\left(\frac{|k^0|}{T},\frac{|\mathbf k|}{T},\pm\frac{\mu_\ell}{T},\pm\frac{\mu_\phi}{T}\right)
\right]
\,,
\\
\hat\Sigma^{{\cal A}i}_{N{\rm L,R}}(k) &= \frac{T^2\hat k^i}{16 \pi |{\mathbf k}|}
\\\notag
&\times
\Bigg[
\vartheta(k^0)
\left(\frac{|k^0|}{|{\mathbf k}|}I_1\left(\frac{|k^0|}{T},\frac{|\mathbf k|}{T},\mp\frac{\mu_\ell}{T},\mp\frac{\mu_\phi}{T}\right) - \frac{M_1^2}{2|{\mathbf k}|T} I_0\left(\frac{|k^0|}{T},\frac{|\mathbf k|}{T},\mp\frac{\mu_\ell}{T},\mp\frac{\mu_\phi}{T}\right) \right)
\\\notag
&-\vartheta(-k^0)
\left(\frac{|k^0|}{|{\mathbf k}|}I_1\left(\frac{|k^0|}{T},\frac{|\mathbf k|}{T},\pm\frac{\mu_\ell}{T},\pm\frac{\mu_\phi}{T}\right) - \frac{M_1^2}{2|{\mathbf k}|T} I_0\left(\frac{|k^0|}{T},\frac{|\mathbf k|}{T},\pm\frac{\mu_\ell}{T},\pm\frac{\mu_\phi}{T}\right) \right)
\Bigg]
\,,
\end{align}
\end{subequations}
where
\begin{align}
I_n(y_0,y,\varrho_\ell,\varrho_\phi) \equiv \int\limits_{\frac{1}{2}(y_0-y)}^{\frac{1}{2}(y_0+y)} dx\;x^n \left(1-\frac{1}{{\rm e}^{x-\varrho_\ell}+1} + \frac{1}{{\rm e}^{y_0-x-\varrho_\phi}-1}\right)\,.
\end{align}
These integrals are given by
\begin{align}
I_0(y_0,y) =&\left[
\log\left({\rm e}^x-{\rm e}^{\varrho_\ell}\right)
-\log\left({\rm e}^{x+\varrho_\phi}-{\rm e}^{y^0}\right)
\right]^{x=\frac12(y^0+y)}_{x=\frac12(y^0-y)}\,,
\\[2mm]\notag
I_1(y_0,y) =& 
\Big[
x\left(
\log\left(1+{\rm e}^{x-\varrho_\ell}\right)
-\log\left(1-{\rm e}^{x-y^0+\varrho_\phi}\right)
\right)
\\\notag
&
+{\rm Li}_2\left(-{\rm e}^{x-\varrho_\ell}\right)
-{\rm Li}_2\left({\rm e}^{x-y^0+\varrho_\phi}\right)
\Big]^{x=\frac12(y^0+y)}_{x=\frac12(y^0-y)}
\,,
\end{align}
where ${\rm Li}_2$ is the dilogarithm.
Of course, for $\mu_\ell=\mu_\phi=0$,
$\hat\Sigma^{{\cal A}\mu}_{N{\rm L}}
=\hat\Sigma^{{\cal A}\mu}_{N{\rm R}}$.

For the scalar model, we do not use the explicit form
of the equilibrium self-energies for the studies
presented in this paper. For completeness, we note that these can be calculated by defining
\begin{align}
p^0_\pm=\frac1{2k^2}
\left\{
k^0 (k^2+m_a^2-m_b^2)
\pm\sqrt{\mathbf k^2(k^2+m_a^2-m_b^2)^2-4\mathbf k^2 k^2 m_a^2}
\right\}
\end{align}
and
\begin{align}
&P(k,G,m_a.m_b,s_a,s_b,L_f)
\\\notag
&\hskip.25cm=G\frac{k^2}{|\mathbf k|}
{\rm sign}^{L_f}(k^2)
\left\{
\vartheta(-k^2)k^0
-p^0_+ +p^0_-
+\frac1\beta\log
\left|
\frac{{\rm e}^{\beta p^0_+}-s_a}{{\rm e}^{\beta p^0_-}-s_a}
\right|
-\frac1\beta\log
\left|
\frac{{\rm e}^{\beta(k^0-p^0_+)}-s_b}{{\rm e}^{\beta(k^0-p^0_-)}-s_b}
\right|
\right\}
\\\notag
&\textnormal{for}\;\;k^2\geq0
\,.
\end{align}
The equilibrium spectral self-energy for the neutral scalar field
$\chi$ is then given by
\begin{align}
\Pi^{\cal A}_{\chi_{ij}}(k)
&=P\left(k,\frac{g_1 g_2^*+g_1^* g_2}{16\pi},m_\varphi,m_\varphi,+1,+1,0\right)
\,.
\end{align}

\section{Green Functions for Leptogenesis}
\label{appendix:propagatorshiggslepton}

The propagators for the leptons are given by
\begin{subequations}
\label{prop:ell:expl}
\begin{align}
{\rm i}S_{\ell}^{<}(p)
&=-2\pi\delta(p^2)P_{\rm L}p\!\!\!/P_{\rm R}\left[
\vartheta(p_0)f_{\ell}(\mathbf{p})
-\vartheta(-p_0)(1-\bar f_{\ell}(-\mathbf{p}))
\right]\,,\\
{\rm i}S_{\ell}^{>}(p)
&=-2\pi\delta(p^2)P_{\rm L}p\!\!\!/P_{\rm R}\left[
-\vartheta(p_0)(1-f_{\ell}(\mathbf{p}))
+\vartheta(-p_0)\bar f_{\ell}(-\mathbf{p})
\right]\,,\\
{\rm i}S_{\ell}^{T}(p)
&=
P_{\rm L}\frac{{\rm i}p\!\!\!/}{p^2+{\rm i}\varepsilon}P_{\rm R}
-2\pi\delta(p^2)P_{\rm L}p\!\!\!/P_{\rm R}\left[
\vartheta(p_0)f_{\ell}(\mathbf{p})
+\vartheta(-p_0)\bar f_{\ell}(-\mathbf{p})
\right]\,,\\
{\rm i}S_{\ell}^{\bar T}(p)
&=
-P_{\rm L}\frac{{\rm i}p\!\!\!/}{p^2-{\rm i}\varepsilon}P_{\rm R}
-2\pi\delta(p^2)P_{\rm L}p\!\!\!/P_{\rm R}\left[
\vartheta(p_0)f_{\ell}(\mathbf{p})
+\vartheta(-p_0)\bar f_{\ell}(-\mathbf{p})
\right]
\,.
\end{align}
\end{subequations}
Due to the unbroken ${\rm SU}(2)_{\rm L}$ symmetry, we suppress the
indices representing the lepton doublet. In principle, the propagators
are diagonal matrices
\begin{align}
S^{{\rm SU}(2)}_{\ell}(u,v)=\delta_{AB}S_{\ell}(u,v)
\,,\qquad A,B=1,2\,.
\end{align}

Whenever these propagators appear within the integrals for the source for
the asymmetry $S$ or the singlet neutrino self-energy $\slashed\Sigma_N$,
it is a sufficient approximation to take for $f_\ell(\mathbf p)$
and $\bar f_\ell(\mathbf p)$ the
equilibrium Fermi-Dirac distribution with zero chemical
potential. Only in the washout term $W$, Eq.~(\ref{washout}),
one should correct this by introducing
a lepton chemical potential, in order to describe a charged distribution
in kinetic equilibrium {\it cf.} Eq.~(\ref{lepto:chempot}).

The propagators for the Higgs fields are
\begin{subequations}
\label{prop:phi:expl}
\begin{align}
{\rm i}\Delta_\phi^<(p)&=
2\pi \delta(p^2)\left[
\vartheta(p_0) f_\phi(\mathbf p)
+\vartheta(-p_0) (1+\bar f_\phi(-\mathbf p))\right]
\,,
\\
{\rm i}\Delta_\phi^>(p)&=
2\pi \delta(p^2)\left[
\vartheta(p_0) (1+f_\phi(\mathbf p))
+\vartheta(-p_0) \bar f_\phi(-\mathbf p)\right]
\,,
\\
{\rm i}\Delta_\phi^T(p)&=
\frac{\rm i}{p^2+{\rm i}\varepsilon}+
2\pi \delta(p^2)\left[
\vartheta(p_0) f_\phi(\mathbf p)
+\vartheta(-p_0) \bar f_\phi(-\mathbf p)\right]
\,,
\\
{\rm i}\Delta_\phi^{\bar T}(p)&=
-\frac{\rm i}{p^2-{\rm i}\varepsilon}+
2\pi \delta(p^2)\left[
\vartheta(p_0) f_\phi(\mathbf p)
+\vartheta(-p_0) \bar f_\phi(-\mathbf p)\right]
\,.
\end{align}
\end{subequations}
As for the leptons, it is understood that
\begin{align}
\Delta^{{\rm SU}(2)}_{\phi  AB}(u,v)=\delta_{AB} \Delta_{\phi}(u,v)\,
\quad A,B=1,2\,.
\end{align}
Again, we can substitute equilibrium distributions in the collision integrals.
When taking account of spectator effects, one should also assign chemical potentials
to the Higgs field and the remaining Standard Model degrees of freedom.

When assuming thermal equilibrium distributions
and vanishing chemical potentials, the Kubo-Martin-Schwinger
(KMS) relations are
\begin{subequations}
\label{KMS}
\begin{align}
{\rm i}S_\ell^>(k)&=-{\rm e}^{k^0/T}{\rm i}S_\ell^<(k)\,,
\\
{\rm i}\Delta_\phi^>(k)&={\rm e}^{k^0/T}{\rm i}\Delta_\phi^<(k)\,.
\end{align}
\end{subequations}
Notice that the according relations also hold for the fermionic and scalar
equilibrium self-energies.

\section{Self-energies from the 2PI Effective Action}
\label{appendix:selferg}
The two-loop contribution 2PI effective action arising from the Yukawa interaction in Eq.~(\ref{Lagrangian}) is given by
\begin{align}
\label{2PI}
\Gamma_2^{(2)} = - g_w Y_i^* Y_j \sum_{a,b=\pm} ab \int d^4x d^4y\,{\rm tr}\left[P_{\rm R} S_{ij}^{ab}(x,y) P_{\rm L}S_{\ell}^{ba}(y,x)\right] \Delta_\phi^{ba}(y,x)\,, 
\end{align}
where $a,b=\pm$ denote the CTP indices, the trace is over Dirac indices and sum over neutrino flavours $i$ is understood. The one-loop neutrino self-energy is then given by 
\begin{align}
\label{Sigma:N1}
{\rm i}\Sigma^{ab}_{Nij}(x,y) =& ab \frac{\delta \Gamma_2}{\delta S^{ba}_{ji}(y,x)}\\
=& g_w \Big(Y_i Y_j^* P_{\rm L} {\rm i}S_{\ell}^{ab}(x,y) P_{\rm R} 
{\rm i} \Delta_\phi^{ab}(x,y)
+ Y_i^* Y_j C P_{\rm R} {\rm i}S_{\ell}^{ba t}(y,x) C^\dagger
P_{\rm L} {\rm i} \Delta_\phi^{ba}(y,x) \Big)\,.
\end{align}
Two contributions arise from the functional derivative, because $S^{ab}$ and $S^{ba}$ are not independent but related by the Majorana symmetry: $S^{ab}_{ij}(x,y) = C S^{ba}_{ji}(y,x) C^\dagger$. Similarly, using $g_w S_{\ell} = \delta_{AB} S^{{\rm SU}(2)}_{\ell  AB}$ in Eq.~(\ref{2PI}) we get for the the lepton self-energy
\begin{align}
\label{Sigma:ell1}
{\rm i}\Sigma^{{\rm SU}(2)\,ab}_{\ell AB}(x,y) =& ab \frac{\delta \Gamma_2}{\delta S^{{\rm SU}(2)\,ba}_{\ell BA}(y,x)}
= Y_i^* Y_j P_{\rm R}
{\rm i}S_{ij}^{ab}(u,v) P_{\rm L}
{\rm i}\Delta_{\phi}^{ba}(v,u) \delta_{AB}
\nonumber\\
\equiv& {\rm i}\Sigma^{ab}_{\ell}(x,y) \delta_{AB}\,.
\end{align}

\section{Explicit Form of the Helicity Block-Diagonal Decomposition}

In order quickly see the relation of
the functions $g_{ah}$ with the vector, scalar, pseudoscalar
and pseudovector densities, it is useful to
recast Eq.~(\ref{helicity:decomposition}) as
\begin{align}
\label{Bloch:explicit}
{\rm i}S_N=\sum\limits_{h=\pm}&-\frac14
\left[
\left(
\begin{array}{cc}
0 & \mathbbm 1\\
\mathbbm 1 & 0
\end{array}
\right)
+h \hat k^i
\left(
\begin{array}{cc}
0 & \sigma^i\\
\sigma^i & 0
\end{array}
\right)
\right]g_{0h}
\\\notag
&
-\frac14
\left[
\left(
\begin{array}{cc}
\mathbbm 1 & 0\\
0 & \mathbbm 1
\end{array}
\right)
+h \hat k^i
\left(
\begin{array}{cc}
\sigma^i & 0\\
0 & \sigma^i
\end{array}
\right)
\right]g_{1h}
\\\notag
&
-\frac14
\left[
\left(
\begin{array}{cc}
{\rm i}\mathbbm 1 & 0\\
0 & -{\rm i}\mathbbm 1
\end{array}
\right)
+h \hat k^i
\left(
\begin{array}{cc}
{\rm i}\sigma^i & 0\\
0 & -{\rm i}\sigma^i
\end{array}
\right)
\right]g_{2h}
\\\notag
&
-\frac14
\left[
\left(
\begin{array}{cc}
0 & -\mathbbm 1\\
\mathbbm 1 & 0
\end{array}
\right)
+h \hat k^i
\left(
\begin{array}{cc}
0 & -\sigma^i\\
\sigma^i & 0
\end{array}
\right)
\right]g_{3h}\,.
\end{align}

\end{appendix}

\end{document}